\documentclass[pdflatex,sn-mathphys-num]{sn-jnl}
\usepackage{graphicx}%
\usepackage{dsfont}
\usepackage{ulem}
\usepackage{multirow}%
\usepackage{amsmath,amssymb,amsfonts}%
\usepackage{amsthm}%
\usepackage{mathrsfs}
\usepackage{calrsfs}%
\usepackage[title]{appendix}%
\usepackage[usenames,dvipsnames,x11names]{xcolor}
\usepackage{textcomp}%
\usepackage{manyfoot}%
\usepackage{booktabs}%
\usepackage{algorithm}%
\usepackage{algorithmicx}%
\usepackage{algpseudocode}%
\usepackage{listings}%
\usepackage{graphicx}
\usepackage{subcaption}
\usepackage{lineno}
\usepackage{pdfpages}
\usepackage{orcidlink}
\setcitestyle{comma,open={},close={},super}
\newcommand*{\citen}[1]{%
  \begingroup
    \romannumeral-`\x % remove space at the beginning of \setcitestyle
    \setcitestyle{numbers}%
    \cite{#1}%
  \endgroup   
}
\newcommand*{\figref}[1]{%
  \hyperref[{#1}]{%
    Fig.~\ref*{#1}%
  }%
}
\renewcommand*{\eqref}[1]{%
  \hyperref[{#1}]{%
    equation~(\ref*{#1})%
  }%
}
\newcommand*{\tabref}[1]{%
  \hyperref[{#1}]{%
    Table~\ref*{#1}%
  }%
}
\begin{document}
%%%%%%%%%%%%%%%%%%%%%%%%%%%%%%%%%%%%%%%%%%%%%%%%
\setlength{\abovedisplayskip}{10pt}
\setlength{\belowdisplayskip}{10pt}
\setlength{\abovedisplayshortskip}{10pt}
\setlength{\belowdisplayshortskip}{10pt} 
%%%%%%%%%%%%%%%%%%%%%%%%%%%%%%%%%%%%%%%%%%%%%%%%
%\linenumbers
%%%%%%%%%%%%%%%%%%%%%%%%%%%%%%%%%%%%%%%%%%%%%%%%
% Title page
%%%%%%%%%%%%%%%%%%%%%%%%%%%%%%%%%%%%%%%%%%%%%%%%
\title[Iterative variational learning of committor-consistent transition pathways using artificial neural networks]
{Iterative variational learning of committor-consistent transition pathways using artificial neural networks}

\author[1,2]{Alberto Meg\'ias \orcidlink{0000-0002-7889-1312}}
\equalcont{These authors contributed equally to this work.}

\author[1]{Sergio Contreras Arredondo \orcidlink{0009-0006-0428-4962}}
\equalcont{These authors contributed equally to this work.}

\author[1]{Cheng Giuseppe Chen \orcidlink{0000-0003-3553-4718}}
\equalcont{These authors contributed equally to this work.}

\author[1]{Chenyu Tang \orcidlink{0000-0002-6914-7348}}
\equalcont{These authors contributed equally to this work.}

\author[4]{Beno\^\i t Roux \orcidlink{0000-0002-5254-2712}}

\author*[1,3,4,5]{Christophe Chipot \orcidlink{0000-0002-9122-1698} }\email{chipot@illinois.edu}

\affil*[1]{Laboratoire International Associ\'e Centre National de la Recherche Scientifique et University of Illinois at Urbana-Champaign, Unit\'e Mixte de Recherche n$^\circ$7019, Universit\'e de Lorraine, B.P. 70239, 54506 Vand\oe uvre-l\`es-Nancy cedex, France}
\affil[2]{Complex Systems Group and Department of Applied Mathematics, Universidad Polit\'ecnica de Madrid, Av. Juan de Herrera 6, E-28040 Madrid, Spain}
\affil[3]{Beckman Institute, 
University of Illinois at Urbana-Champaign, Urbana, USA}
\affil[4]{Department of Biochemistry and Molecular Biology, 
University of Illinois at Urbana-Champaign, Urbana, USA}
\affil[5]{Department of Physics, University of Chicago, Chicago, USA}

%%%%%%%%%%%%%%%%%%%%%%%%%%%%%%%%%%%%%%%%%%%%%%%%
% Abstract
%%%%%%%%%%%%%%%%%%%%%%%%%%%%%%%%%%%%%%%%%%%%%%%%

\abstract{This contribution introduces a neural-network-based approach to discover meaningful transition pathways underlying complex biomolecular transformations in coherence with the committor function. The proposed path-committor-consistent artificial neural network (PCCANN) iteratively refines the transition pathway by aligning it to the gradient of the committor. This method addresses the challenges of sampling in molecular dynamics simulations rare events in high-dimensional spaces, which is often limited computationally. Applied to various benchmark potentials and biological processes such as peptide isomerization and protein-model folding, PCCANN successfully reproduces established dynamics and rate constants, while revealing bifurcations and alternate pathways. By enabling precise estimation of transition states and free-energy barriers, this approach provides a robust framework for enhanced-sampling simulations of rare events in complex biomolecular systems.}


%%%%%%%%%%%%%%%%%%%%%%%%%%%%%%%%%%%%%%%%%%%%%%%%
\maketitle
%%%%%%%%%%%%%%%%%%%%%%%%%%%%%%%%%%%%%%%%%%%%%%%%

Characterizing the conformational changes underlying the function of biological macromolecules is a longstanding computational problem of paramount importance. Considering the rare conformational transitions between two long-lived metastable states, $A$ and $B$, as seen by molecular dynamics (MD) simulations, theoretical and computational frameworks have traditionally been formulated around the concept of a reaction coordinate between the two states\cite{Eyring-1935,Wigner-1938,Laidler-1983}. In time, these ideas were rigorously cast in the language of statistical mechanics\cite{Chandler-rate,Berne-rare,Berne-1988}, leading to the notion of a transition-path ensemble\cite{bolhuis_transition_2000}. Strictly speaking, the latter must be characterized by sampling the space of trajectories with unbiased simulations\cite{bolhuis_transition_2000}.  

However, many of the conformational transitions underlying complex biological processes must overcome significant free-energy barriers, making it difficult, if not impossible, to sample the transition-path ensemble with unbiased MD trajectories using common computational resources. The string method was formulated to bridge the classic notion of reaction coordinate with the complex reality of the transition-path ensemble by providing a ``reaction tube'' encompassing the dominant reactive trajectories\cite{E2002string,e_finite_2005,maragliano2006string,pan_finding_2008}. In practice, the string pathway is represented as a set of discretized intermediate configurations, or images, connecting end-states $A$ and $B$ in a subspace of collective variables (CVs). The latter are mathematical funct10ions combining important degrees of freedom that must be chosen carefully to render meaningful dynamics\cite{rogal_reaction_2021}. The string provides a convenient framework to generate biased simulations for sampling the path-collective variables (PCVs),\cite{branduardi_b_2007}and infer important thermodynamic and kinetic properties such as free-energy barriers and rate constants\cite{hollingsworth_molecular_2018}.

Nevertheless, despite its appealing features, the string\cite{E2002string,e_finite_2005,maragliano2006string,pan_finding_2008} is not necessarily consistent with the true ensemble of reactive transition paths by construction. An important ingredient to shed light on this issue is provided by the committor, defined as the probability that a trajectory initiated at some random configuration will ultimately reach state $B$ before state $A$.\cite{Onsager-1938} It has been shown, in the context of transition-path theory (TPT), to be a critical feature associated with the true reaction coordinate controlling the transitions in the system\cite{bolhuis_reaction_2000,Vanden-Eijnden-2010,berezhkovskii_committors_2019,roux_string_2021,roux2022transition} . Thus, it is crucial that any path-finding algorithm be aware of the underlying committor to properly describe the underlying dynamics. Said differently, the simultaneous requirement to identify the correct reaction pathway from adequate sampling is akin to a chicken-and-egg problem, suggesting that one must consider an iterative approach, whereby sampling is performed along a guess pathway employing biased simulations from which a new, better path is inferred. 

In this contribution, we put forth such an iterative approach to refine the reaction path between two metastable states following from the concepts of the committor-consistent variational string method\cite{he_committor-consistent_2022} and the variational committor network (VCN), which determines the committor from biased simulations\cite{chen_discovering_2023}. In a nutshell, the variational principle is invoked to minimize a two-point time-correlation function of the committor, $q$,
\begin{equation}\label{eq:Cqq}
C_{qq}(\tau)= \frac{1}{2}\left\langle (q(\tau)-q(0))^2\right\rangle,
\end{equation}
\noindent where $\tau$ is the time lag required to render the dynamics Markovian in the subspace of the CVs.\cite{roux_string_2021,roux2022transition}

The present work has similarities with recent efforts \cite{khoo2019solving, rotskoff2022active,chen2023committor, jung2023machine, kang2024Parrinello} applying various artificial neural networks (ANNs) to learn the committor function. Modern path-finding and dimensionality-reduction schemes, such as ANNs\cite{goodfellow_deep_2016}, aim at identifying the critical CVs from a set of candidate coordinates\cite{chen_collective_2018,brandt_machine_2018,mardt_vampnets_2018,wang_pastfuture_2019,chen_nonlinear_2019,rydzewski_multiscale_2021,odstrcil_lines_2022,sipka_differentiable_2023, Bonati_2023, FranceLanord2024, bonati2021deep, yang2024learning}. These ideas may find applications that go far beyond understanding chemical and biological processes, for instance in climate \cite{lucente2022climate} and ocean studies\cite{jacques2023ocean}. However, while some research efforts have considered the variational principle in the context of an overdamped diffusive dynamics approximation\cite{khoo2019solving,kang2024Parrinello}, these schemes are only valid in the limit of an infinitesimal time lag ($\tau \rightarrow 0^+$), and are, hence, system-dependent\cite{roux_string_2021,roux2022transition}. The premises of \eqref{eq:Cqq} with a finite time lag are, however, more general\cite{he_committor-consistent_2022,chen_discovering_2023}.

\begin{figure}[ht!]
	\centering
   \includegraphics[width=1.0\textwidth]{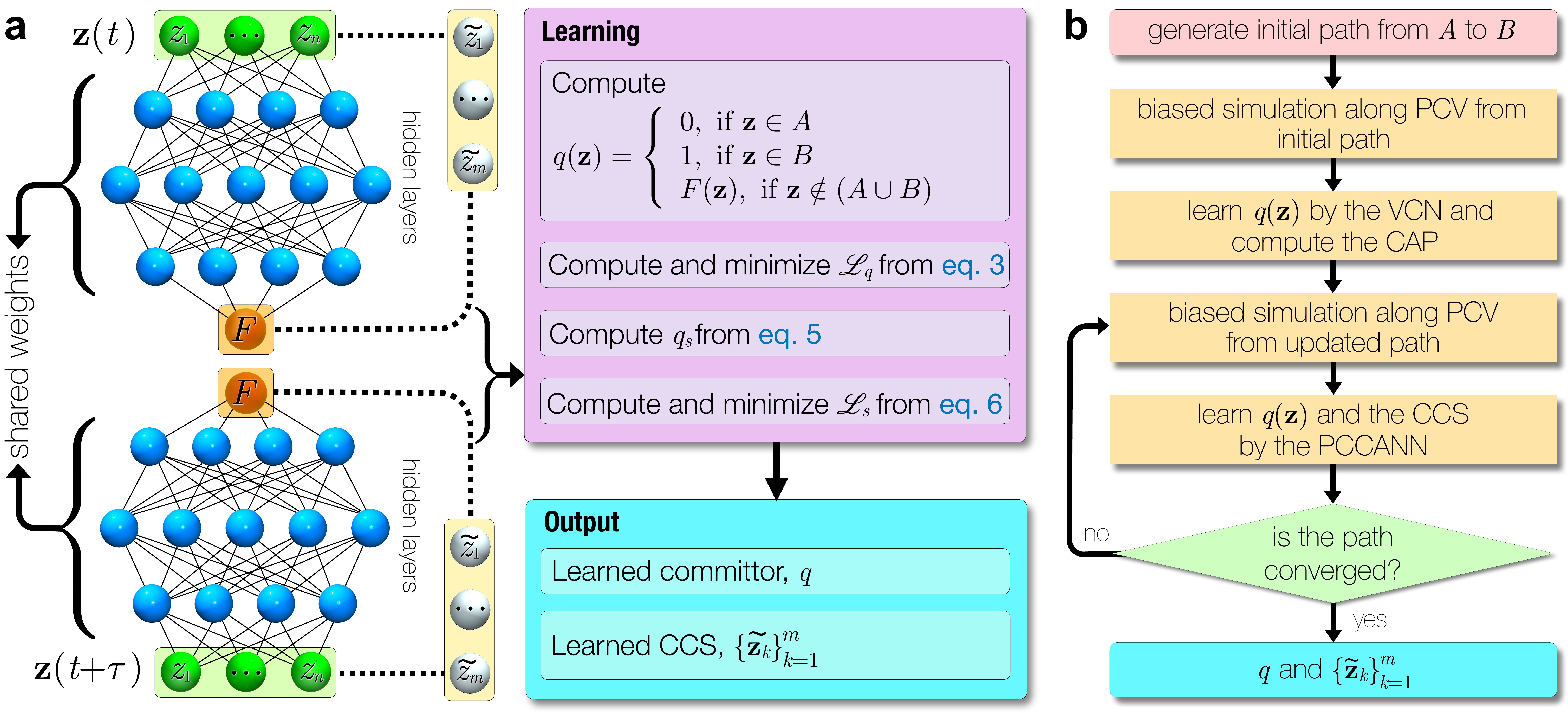}
	\caption{\label{fig:vcns_iterative} Schematic representation of (a) the proposed ANN for learning simultaneously the committor and the CCS and (b) the workflow to find a final committor and CCS.}
\end{figure}

Our proposed iterative approach follows the spirit of the VCN, and rests on a new architecture coined path-committor-consistent ANN (PCCANN). This approach consists of (i) learning a committor function from biased trajectories of candidate CVs through a variational approach using ANNs, (ii) aligning the initial path to a committor-consistent one, referred, henceforth, to as committor-consistent string (CCS), and (iii) generating new trajectories from biased MD along the updated path. The extension of the VCN to learn simultaneously the CCS and the committor, together with the iterative procedure are summarized in \figref{fig:vcns_iterative}. In \nameref{sec:method}, we develop the theoretical foundations of their design.

%%%%%%%%%%%%%%%%%%%%%%%%%%%%%%%%%%%%%%%%%%%%%%%%
\section*{Results}\label{sec:Results}
%%%%%%%%%%%%%%%%%%%%%%%%%%%%%%%%%%%%%%%%%%%%%%%%
The performance of the PCCANN is examined in a series of paradigmatic illustrations. As a proof of concept, a one-dimensional double-well potential is used to recover the committor, as detailed in the Supplementary Information (SI). Next, our iterative approach is probed in problems of higher dimensionality, starting with the two-dimensional Berezhkovskii--Szabo\cite{berezhkovskii_one-dimensional_2005} (BS) and M\"uller--Brown\cite{Mller1979LocationOS} (MB) analytical potentials. The PCCANN was also applied to the prototypical $N$--acetyl--$N$$^\prime$--methylalanylamide (NANMA) as a preamble to processes characterized by multiple pathways, namely a two-dimensional triple-well (TW) potential\cite{metzner_2008}, and the artificial mini-protein chignolin, composed of ten amino acids (GYDPETGTWG). For all biased MD simulations, the well-tempered meta-extended adaptive biasing force (WTM-eABF) algorithm\cite{fu_taming_2019} was employed along a PCV.

\subsection*{Berezhkovskii--Szabo Potential}

We looked for the CCS between the two metastable states of the BS potential\cite{berezhkovskii_one-dimensional_2005} using the PCCANN. Although this potential is analytical and does not present any inherent sampling difficulty, the committor must be mapped accurately to reflect the anisotropic diffusivities ($D_x\neq D_y$), or drag coefficients ($\gamma_x\neq \gamma_y$). This challenge has motivated a number of other studies\cite{Tiwary2017Predicting,he_committor-consistent_2022,chen_discovering_2023}. We have generated unbiased and biased trajectories for different ratios of $\gamma_x/\gamma_y$ to compare the isotropic ($\gamma_x/\gamma_y = 1.0$) and anisotropic diffusivity cases ($\gamma_x/\gamma_y = 10.0$ corresponds to a higher diffusivity in the $y$--direction, and $\gamma_x/\gamma_y = 0.1$, to a higher diffusivity in the $x$--direction). 

\begin{figure}[ht!]
\centering
\includegraphics[width=\textwidth]{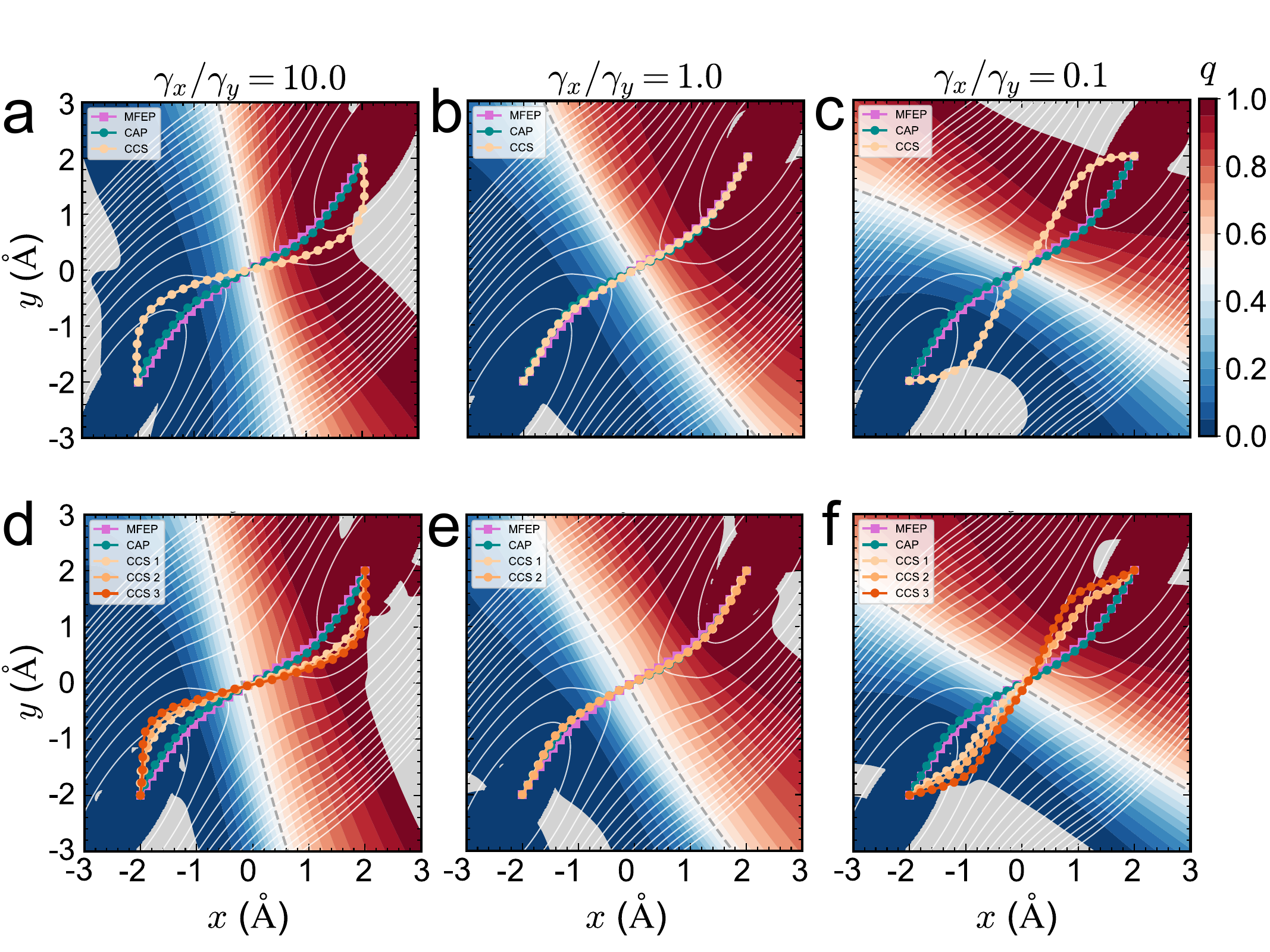}
	\caption{\label{fig:BS-results} Learned committor function, $q(x,y)$, for three unbiased (a--c) and biased (d--f) trajectories for the BS potential. The contour lines in white show the potential-energy surface. The gray dashed line highlights the separatrix. The line with squares denotes the MFEP, and those with circles, the different iterations for the CCS. The left (a, d) and right (c, f) columns depict the anisotropic diffusion cases with $\gamma_x/\gamma_y=0.1$ and $\gamma_x/\gamma_y=10.0$, respectively, whilst the center one (b, e) represents the isotropic case.}
\end{figure}

The PCCANN results for the BS potential are gathered in \figref{fig:BS-results}. In the first row (\figref{fig:BS-results}a--c), it is apparent that the PCCANN perfectly captures the effect of the anisotropic diffusivity, and the symmetry inherent to the form of the potential is preserved. Neither the reference minimum free-energy pathway (MFEP) obtained using the Dijkstra algorithm\cite{fu2020finding}, nor the committor-averaged path (CAP, see \nameref{sec:method}), computed from the committor learned by the VCN and closely aligned with the MFEP, render the asymmetry in the diffusion that the CCS perfectly reflects.

In \figref{fig:BS-results}d--f, we followed the iterative procedure detailed in \nameref{subsec:iterative}. For the anisotropic cases, at least three iterations were needed to recover the results obtained in the unbiased case, whereas for the isotropic case, only one iteration was necessary. The CCSs determined at each iteration reflect both the asymmetric diffusion and the symmetry inherent to the potential. As expected, the CCSs from both biased and unbiased MD cross the saddle point, perpendicularly to the $q=1/2$ isocommittor plane, also known as the separatrix. This criterion represents the best choice to determine the direction of a string that defines an effective PCV, i.e., one-dimensional reaction coordinate, as has been discussed recently\cite{roux2022transition,he_committor-consistent_2022}.
Finally, the rate constants are inferred from the converged committor for both unbiased and biased simulations, and agree with the results derived from the mean-first-passage time (MFPT) (see SI).

\subsection*{M{\"u}ller-Brown Potential}

Another well-known model system is the MB potential\cite{Mller1979LocationOS}. This potential is particularly challenging for optimization algorithms due to its complex landscape, which includes multiple local minima and saddle points. These features render difficult the discovery of the global minimum, and, by extension, transition states. As a result, the MB potential has become a standard benchmark for testing and evaluating new algorithms\cite{NOE_2017,Bonati_2023,FranceLanord2024,Muhammad_2022}.

\begin{figure}[ht!]
\centering
\includegraphics[width=\textwidth]{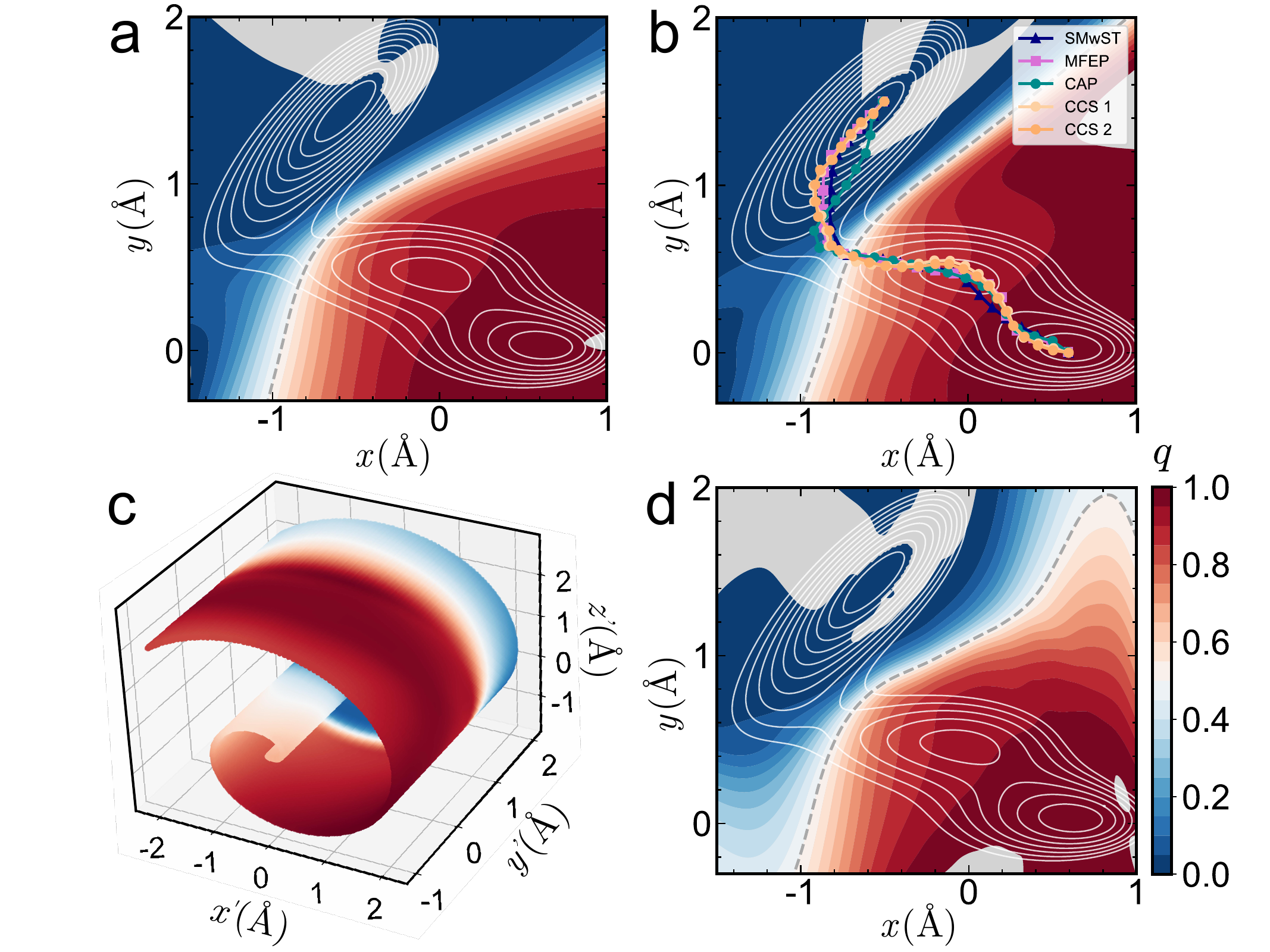}
    \caption{\label{fig:mb_ccs} Learned committor function, $q(x,y)$, for unbiased (a) and biased (b) trajectories for the MB potential. The contour lines in white show the potential energy surface. The gray dashed line highlights the separatrix. The line with triangles depicts the SMwST, the one with squares, the MFEP, and those with circles, the different iterations for the CCS. The lower-left hand panel (c) represents the committor learned in three dimensions, while panel (d) shows the two-dimensional back-projection.}
\end{figure}

Learning starts with the CAP as the initial guess path, depicted as the green string in \figref{fig:mb_ccs}b, and computed from the VCN using a PCV simulation along a straight path connecting the two metastables states $(x, y) = (-0.6, 1.4)$ and $(x, y) = (0.6, 0.0)$. This CAP provides a better initial path for concomitantly learning the committor and the CCS than a straight path connecting the two basins. The learned committor is shown in \figref{fig:mb_ccs}. Our iterative approach successfully captures the correct dynamics, as reflected by the overall committor, with the converged string crossing the saddle point, and passing as expected through the intermediate local minimum $(x, y) = (-0.1, 0.5)$. Furthermore, the learned path converged within two iterations, closely matching both the MFEP and the path from the SMwST. To asses the effectiveness of the PCCANN compared to the VCN\cite{chen_discovering_2023}, we also learned a committor from unbiased MD, as depicted in \figref{fig:mb_ccs}a. Clearly, the committor generated using biased MD along the converged path closely resembles that produced by unbiased MD. Last, to address one of the greatest challenges posited by complex systems, where the surface upon which the dynamics occurs is usually unknown, we mapped the MB potential onto a Swiss-roll surface, now expressed in a three-dimensional coordinate system (see \figref{fig:mb_ccs}c)\cite{strahan2023predicting}. Details of this mapping is supplied in the SI. Back-projection in two dimensions of the committor learned in three dimensions is depicted in \figref{fig:mb_ccs}d. The committor accurately reflects the correct dynamics near the saddle point, but differs in the higher-energy region where sampling is nearly nonexistent. 

\subsection*{Triple-Well Potential}\label{ssec:triple-well}

The simple TW potential is particularly helpful to illustrate the occurrence of multiple pathways in studying transitions between metastable states. Notably, the TW potential exhibits the phenomenon of ``entropic switching''\cite{metzner_2008}, whereby certain transition paths become dominant depending on the thermal energy of the system. At low temperatures, transitions predominantly occur through an upper channel, while at higher temperatures, transitions favor a lower channel. The importance of TW potentials lies in their ability to highlight the influence of different reaction pathways, entropy, and variations in transition rates as a function of the temperature.

\begin{figure}[ht!]
\centering
\includegraphics[width=0.7\textwidth]{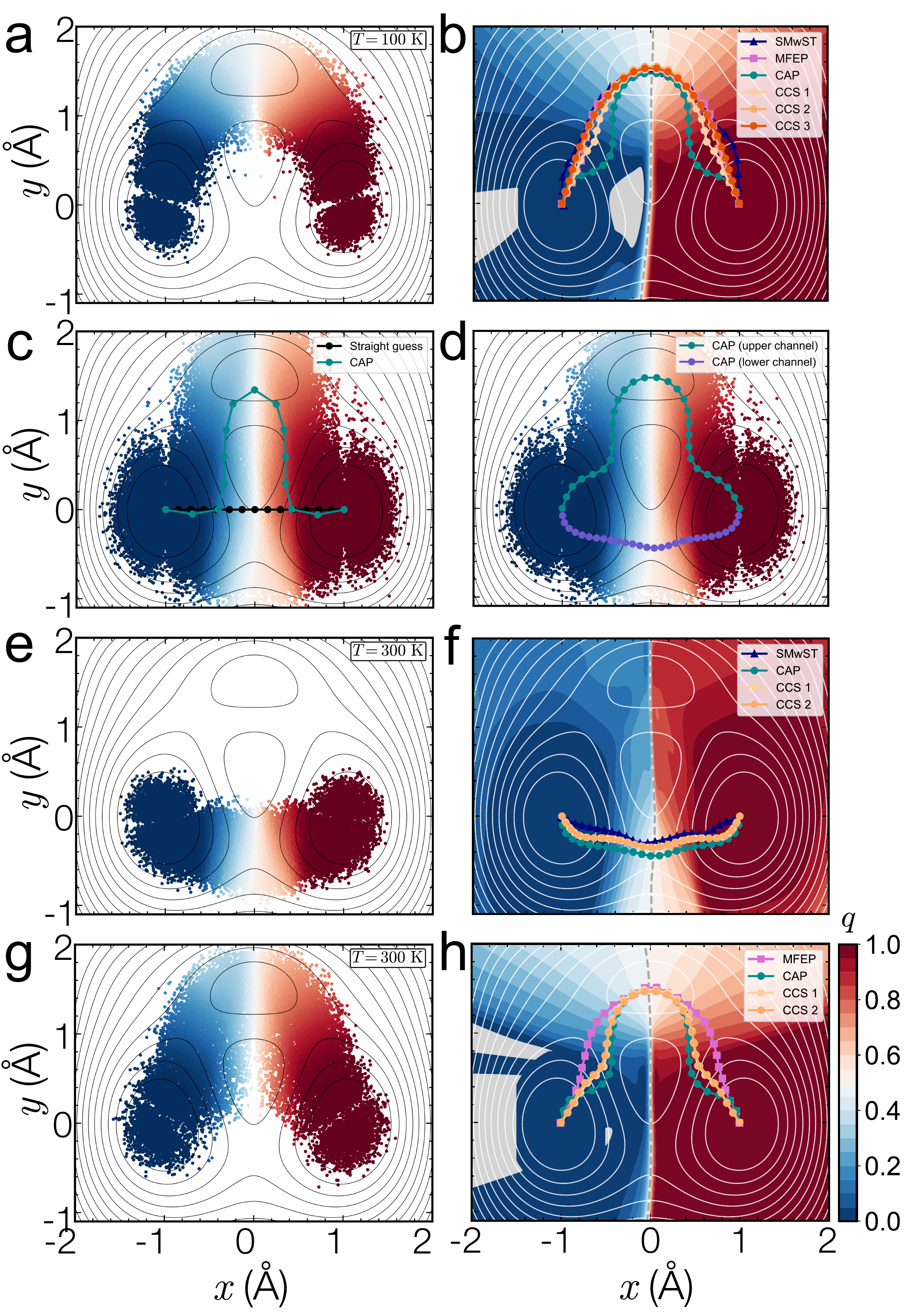}
    \caption{\label{fig:TW} Sampled region from biased MD simulations for the TW potential at 100 K along the converged CCS (a), and 300 K along the straight path (c) and along the CCSs (e,g). The corresponding learned committor function, $q(x,y)$, at 100 K (b) and 300 K (f,h), along with the CCS for each iteration. Averaging the  points over isocommittor slices covering the entire sampled CV space as a way to infer an initial path for the learning process yields a suboptimal CAP (c). The improved CAPs are determined by averaging the points in each isocommittor slice independently for the two reactive tubes identified by clustering (d). The contour lines represent the potential energy surface, while the gray dashed line highlights the separatrix. The lines with triangles represents the SMwST, the ones with squares, the MFEP, and those with circles, the different iterations for the CCS.} 
\end{figure}

To analyze the effect of temperature on the reactive trajectories, biased MD simulations were performed at 100 and 300 K (see SI). At low temperatures, the particle struggles to overcome the potential barrier, and preferentially transitions through the upper channel. When the temperature is sufficiently high, the distribution is primarily located around the deep minima of the potential, displaying metastability, i.e., the particle makes frequent transitions between these minima. The phenomenon of entropic switching, whereby the transition pathway changes with temperature, can be understood within the framework of TPT\cite{metzner_2006}. Sampling confirms that at higher temperatures (\figref{fig:TW}e), the reaction predominantly occurs via the lower channel, while at lower temperatures (\figref{fig:TW}a), it favors the upper channel. This behavior aligns with the findings reported in references \citen{metzner_2006,Schulten_2003,Liang_2023}. 

Our primary objective is to describe the committor and determine the reaction pathway for each channel using the iterative approach developed here. \figref{fig:TW}b, f, and h illustrate the latter, starting from the CAP, from which new paths are successively learned. Owing to a bifurcation of the reactive tube, the concept of CAP is suboptimal, because the committor is averaged over both pathways, as shown in \figref{fig:TW}c. 
Alternatively, a more sophisticated strategy based on a clustering analysis of committor-sorted configurations extracted from the simulation is able to identify two distinct transition paths. The result is shown in Fig. S3 of the SI, where the two reaction paths can clearly be identified by the committor-based K-means clustering\cite{lloyd1982kmeans}, from which the initial strings can be subsequently determined from the centroids of the clusters. For the sake of clarity, the configurations extracted from the simulation were simply sorted according to the value of the $y$ coordinate, $y>0$ and $y<0$ being ascribed to the upper and lower transition path, respectively, in order to obtain reasonable estimates of the CAPs for the two paths. The estimated CAP for the upper and lower transition paths determined from this simplified procedure are shown in \figref{fig:TW}d. The result is consistent with the clustering analysis shown in Fig. S3. Applying this method to the TW potential yields improved guess paths, as shown in \figref{fig:TW}b, f, and h. At low temperature (\figref{fig:TW}b), we observe that the path evolves towards the shallow minimum, approaching both the SMwST path and the MFEP, while crossing the saddle points. Conversely, at higher temperature (see \figref{fig:TW}f), most of the reactive flux follows the lower channel, as reflected in the rate constants (see SI), and the CCS, transitioning through the maximum of the potential,  eventually approaches the SMwST path. Moreover, sampling the upper channel by means of confinement potentials yields a CCS (see \figref{fig:TW}h) somewhat different from that observed at low temperature (see \figref{fig:TW}b), because the reactive flux is lower\cite{metzner_2006}, compared to that at 100 K, as confirmed by the associated rate constants (see SI).

\subsection*{$N$--Acetyl--$N^\prime$--methylalanylamide}

\begin{figure}[ht!]
    \centering
    \includegraphics[width=1.0\linewidth]{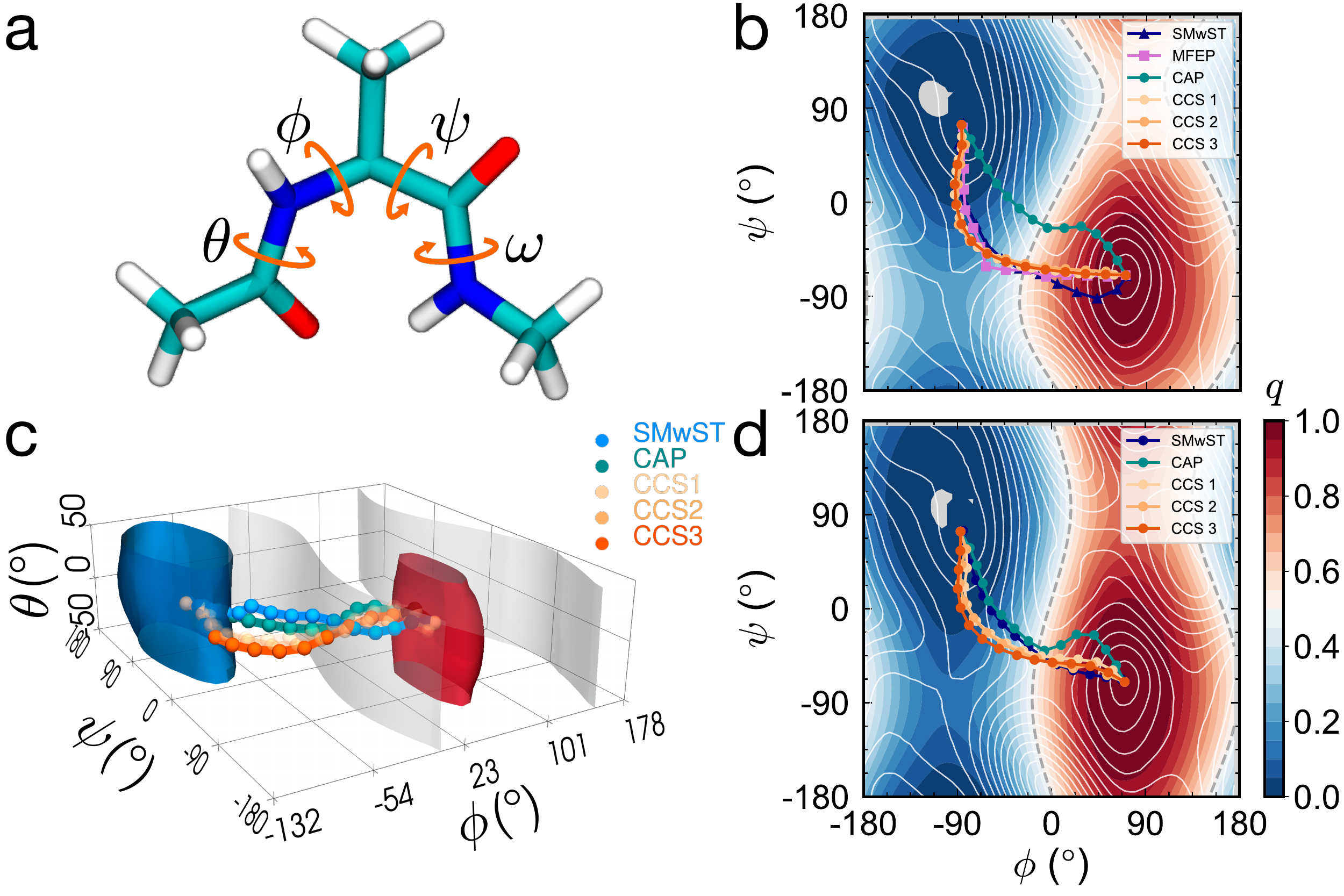}
    \caption{NANMA and the dihedral angles $\phi$, $\psi$, $\theta$ and $\omega$ (a). Learned committor function, $q$, and iterated CCSs in the $(\phi, \psi)$ subspace. The MFEP (in violet) and the path obtained with the SMwST (blue) are also reported as references (b). Learned committor, $q$, and iterated CCSs in the $(\phi, \psi, \theta, \omega)$ subspace: Three-dimensional plot with $\omega=0^\circ$ (c), and two-dimensional plot with $\theta=0^\circ$ and $\omega=0^\circ$ (d). The contour lines represent the free-energy surface, while the gray dashed line highlights the separatrix. The lines with triangles represents the SMwST, the ones with squares, the MFEP, and those with circles, the different iterations for the CCS.}
    \label{fig:nanma}
\end{figure}

We now consider a system of biological relevance, NANMA in vacuum. The isomerization between the C$\mathrm{_{7eq}}$ and C$\mathrm{_{7ax}}$ conformations was tackled previously using the VCN,\cite{chen_discovering_2023} choosing the backbone dihedral angles, $\phi$ and $\psi$, as the features of the ANN. Here, we expand the scope to learn the transition pathway by considering not only the $(\phi,\psi)$ subspace, but also the ancillary $\theta$ and $\omega$ dihedral angles (see \figref{fig:nanma}a). In fact, although commonly overlooked, the importance of $\theta$ to provide a complete description of the conformational equilibrium has been long established.\cite{bolhuis_reaction_2000} We, therefore, trained our model first by using $(\phi,\psi)$, and then $(\phi,\psi,\theta,\omega)$, to compare the effectiveness of our methodology when evolving towards higher-dimensionality models.

Since we are imposing the tangent of the CCS to be parallel to the gradient of the committor, when dealing with more complex systems, the string may be susceptible to getting stuck in a local minimum of the loss function, $\mathcal{L}_\mathrm{s}$ (see \eqref{eq:loss_s}), bearing no physical significance (see SI). In order to avoid this pitfall, multiple models were trained for each iteration, and the one with the lowest value of $\mathcal{L}_\mathrm{s}$ was kept, in accordance to the variational principle. This issue is even more pronounced when increasing dimensionality, which further augments the number of local minima. Consequently, when the $(\phi,\psi,\theta,\omega)$ subspace was considered, we used a modified restraint, $\mathcal{L}^*_\mathrm{tr}$ (see SI), which assigns a lower weight to those contributions where $\|\nabla q(\mathbf{z})\|$ is low, i.e., where the noise is also higher. This precaution was obviated for analytical, less realistic systems, or when the dimensionality of the process at hand was lower, i.e., two or fewer CVs, since the committor behavior was more easily captured (see SI).

Our results are gathered in \figref{fig:nanma}. In the two-dimensional case, we obtain a committor map that is consistent with the one learned by the VCN,\cite{chen2023committor} and a CCS closely resembling the MFEP obtained using the Dijkstra algorithm\cite{fu2020finding} (see \figref{fig:nanma}b). Some discrepancies can be found when comparing the CCS to the SMwST pathway, albeit the rate constants computed using the two strings are very close (see SI), suggestive that the trajectories are contained in similar tubes. When $\theta$ and $\omega$ are also included in the input features, the PCCANN correctly confirms the importance of $\theta$ to characterize the transition path (see \figref{fig:nanma}c and SI). Moreover, the addition of $\omega$, which is less relevant for describing the transition, did not affect the learning of the path in the other dimensions, underscoring the ability of the PCCANN to address so-called nuisance variables. The separatrix plotted in the $(\phi,\theta)$ subspace exhibits the same anticorrelation between $\phi$ and $\theta$ highlighted in previous studies\cite{bolhuis_reaction_2000, maragliano2006string, kang2024Parrinello}. While the CCS and the SMwST (as well as the string reported in reference \citen{maragliano2006string}) are similar when projected onto $(\phi,\psi)$, it is worth noting the discrepancy of the values of $\theta$, as shown in \figref{fig:nanma} and in the SI. Despite showing qualitatively opposite trends along $\theta$, the calculated rate constants are nearly identical, further supporting our results. This coincidence stems from the largely overlapping transitions sampled in the reaction tubes defined by the CCS and the SMwST string (see SI).

\subsection*{Chignolin}\label{ssec:chignolin}

\begin{figure}[ht!]
    \centering
    \includegraphics[width=\linewidth]{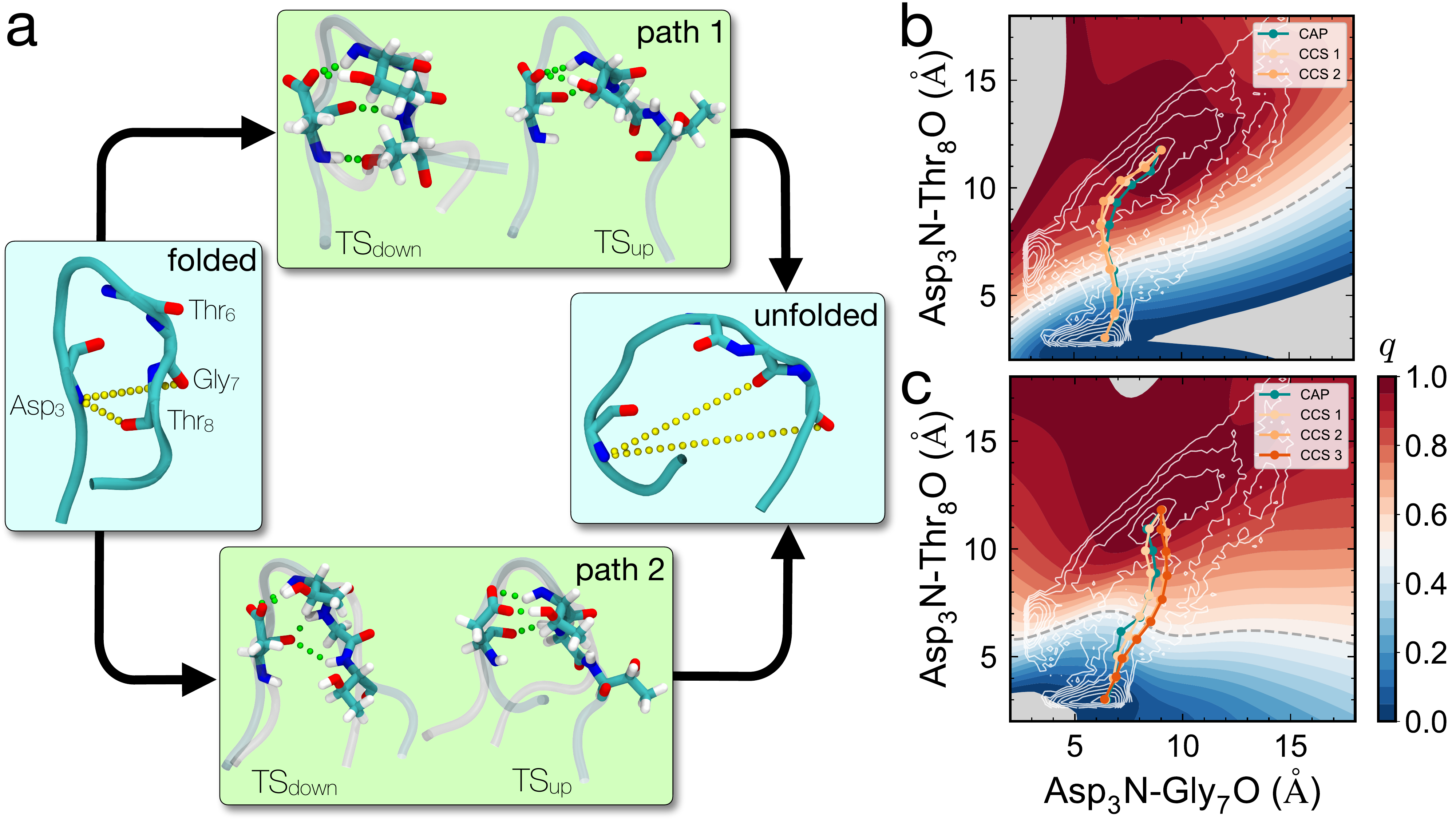}
    \caption{Chignolin in its folded, unfolded and transition states (a). The transition states were obtained by performing a K-medoids clustering\cite{Park2009kmedoids}. Learned committor function, $q$, and CCSs at each iteration in the subspace defined by (Asp$_3$N-Gly$_7$O, Asp$_3$N-Thr$_8$O) for the two folding pathways of chignolin (b,c).
    }
    \label{fig:chignolin}
\end{figure}

As a final example, we applied our iterative method to the study of the reversible folding of the mini-protein chignolin in aqueous solution, which transitions between its native, $\beta$-hairpin conformation, its unfolded state, and a misfolded state. Each of these states is identifiable through a set of two hydrogen bonds\cite{Oshima2019replica, Chen2021} (see also \figref{fig:chignolin}a), which we chose as the CVs for the learning process, and to build our PCV. Owing to the complexity of the system, it would be desirable to consider a larger number of CVs, which is in principle amenable to our approach, as was done for NANMA. In practice, however, considering a large number of CVs presents challenges rooted in the PCV simulation itself---not in the learning. We, therefore, remained in the two-dimensional CV space,\cite{Oshima2019replica, Chen2021} to showcase the strengths of the PCCANN. Chignolin serves also as an interesting benchmark for the methodology implemented for the triple-well potential to handle multiple reaction tubes. 
Two transition states were identified for this folding process, in line with a recent study\cite{kang2024Parrinello} (see \figref{fig:chignolin}). By performing a K--medoids clustering analysis\cite{Park2009kmedoids} using the backbone N---O interatomic distances, we have identified seven relevant conformations representative of the transition states---three belonging to the fist path, and four to the second one (see SI). Upon further inspection, we observed that, in both pathways, the medoids can be classified into two groups, namely clusters that form hydrogen bonds between Asp$_3$ and of Thr$_6$ (either through the side chains or the backbone heteroatoms), and clusters where additional interactions between Asp$_3$ and Thr$_8$ form (see \figref{fig:chignolin}a and SI). These findings also agree with the same recent study\cite{kang2024Parrinello}, and the free-energy difference between the end-states are consistent with the data reported using alternate approaches\cite{yang2024learning,Oshima2019replica,Chen2021}.

%\radu{\subsection*{Isobutanol isomerization}\label{ssec:Isobutane isomerization}
%To further show the expandability of the method beyond force-field based approaches, we investigated the isomerization of isobutanol using ab-initio molecular dynamics, relying on the GFN2-xTB semi-empirical method to speed up sampling. The chemical reaction, characterized in  https://doi.org/10.1021/acscatal.4c00736 (will add as a proper reference if we move forward with this) captures the transformation of isobutanol into butan-2-ol. We show that the free energy profile qualitatively matches previously reported work. The generated string is compared to that obtained from a nudged elastic band approach (calculated on the electronic energy surface) and a SMwST (currently running) optimized string. We can see from the molecular structures that we are able to locate a transition state which matches/differs those delivered from the other string calculation methods (will add information how depending on results). This shows the high transferability of the method and the potential applications to quantum mechanical(QM) and QM/MM reaction mechanism investigations. }

\section*{Discussion}\label{sec12}

In this contribution, we have developed a neural-network-based method for simultaneously learning the committor and its CCS, which has proven highly effective across a range of biophysical systems, demonstrating its ability to reproduce both expected and novel results in systems with varying symmetries and complexities. The PCCANN efficiently infers the impact of anisotropic diffusivities, as demonstrated for the BS potential, where the CCS diverges from the MFEP. In rugged landscapes like the MB potential, our approach successfully captured the correct dynamics and converged to a string that matched both the MFEP and the SMwST pathway. Similarly, for the TW potential, the model was able to trace the alternate reactive tubes as a function of the temperature. The method was also applied to  molecular processes of biological relevance, like NANMA isomerization and chignolin reversible folding. In the former case, the PCCANN yielded results in agreement with the literature\cite{bolhuis_reaction_2000,kang2024Parrinello}, reproducing the anticorrelation between dihedral angles at the transition state. In the latter case, two possible pathways were identified, revealing that folding does not necessarily pass through an intermediate misfolded state\cite{yang2024learning}. Moreover, our estimate of the free-energy difference between the folded and unfolded states aligns with previous findings\cite{Oshima2019replica,Chen2021,kang2024Parrinello,Lindorff-Larsen2011fold}. The proposed method is shown to provide in most cases accurate estimates of the rate constants, consistent with previously reported values (see SI). Put together, the PCCANN offers a promising paradigm for enhanced-sampling simulations, most notably for more complex systems. Nevertheless, the method depends on a careful choice of CVs to featurize the ANN and reproduce the underlying dynamics, which even with highly efficient importance-sampling algorithms like WTM-eABF, remains for certain challenging molecular systems a daunting task. Ongoing efforts focus on refining the loss-function parameters and the unbiasing algorithms to further improve accuracy, aiming to extend the applicability of the method to increasingly intricate systems.

\section*{Methods}\label{sec:method}

In this section, we recap the theoretical foundations of the VCN introduced to learn the committor\cite{chen_discovering_2023}, and outline our novel approach---the PCCANN, designed to learn the CCS, detailing how the latter is related to the committor. We also describe the protocol associated to the iterative procedure that supplies the relevant CCS for the biomolecular process at hand.

\subsection*{Variational Committor Networks}

Let us consider a molecular system of interest with Cartesian coordinates $\mathbf{x} \equiv \lbrace \mathbf{x}_i \rbrace_{i=1}^N$, where $N$ is the number of atoms, and ${\bf x}_i$, the position of $i$-th atom. At temperature $T$, the system is characterized by the Boltzmann equilibrium probability density $\rho_{\rm eq}({\bf x})=e^{-U({\bf x})/T}/Z_T$, where $Z_T$ is the configurational partition function. 
We choose a set of candidate CVs represented as functions of the Cartersian coordinates, $\mathbf{z}(\mathbf{x}) \equiv \lbrace z_j(\mathbf{x}) \rbrace_{j=1}^n$, where $n\leq N$ is the number of CVs. 
% DO WE NEED THIS DEFINITION???
%The marginal probability distribution in the space of the CVs can be expressed as $\rho({\bf z})= \int d \mathbf{x} \, \delta({\bf z}({\bf x})-{\bf z}) \, \rho_{\rm eq}({\bf x})$.  
We assume that the molecular system comprises two basins, $A$ and $B$. Accordingly, one defines the forward committor at a given value of the CVs, $q(\mathbf{z})$, as the sum of the probabilities over all paths starting at a given reference position in CV space, $\mathbf{z}$, that reach state $B$ before crossing state $A$. The probability of each path can be obtained as a product of discrete propagation steps with time lag $\tau$, under the restriction that the intermediate states of these paths are neither $A$ nor $B$\cite{roux_string_2021}.
The committor can be determined by a variational principle based on minimizing either the reaction flux, $J_{AB}$, or, equivalently, the two-point correlation function, $C_{qq}$\cite{roux_string_2021,he_committor-consistent_2022,chen_discovering_2023}.
In the context of TPT, the net forward reactive flux 
from $A$ to $B$ is defined as $J_{AB}=C_{qq}(\tau)/\tau$, with $C_{qq}$ given by \eqref{eq:Cqq}. This quantity can be computed numerically from a simulation by splitting a trajectory in shorter trajectories of length $\tau$\cite{roux_string_2021}.

The underlying idea of the present work is based on references~\citen{he_committor-consistent_2022} and \citen{chen_discovering_2023}, 
where the committor function is approximated using a generic $\mathbf{z}$-dependent scalar function constrained by $q(\mathbf{z})=0$ for all $\mathbf{z} \in A$, and $q(\mathbf{z})=1$ for all $\mathbf{z} \in B$, and $q(\mathbf{z})$ takes intermediate values elsewhere, as depicted in \figref{fig:vcns_iterative}.
According to the universal approximation theorem~\cite{cybenko_approximation_1989}, ANNs are good choices for function approximation. Therefore, as shown in \figref{fig:vcns_iterative}, $q(\mathbf{z})$ is approximated by the ANN function $F(\mathbf{z})$. In practice, the Siamese ANN~\cite{bromley_signature_1993,chicco_siamese_2021} architecture is used for processing the inputs ${\bf z}(t)$ and ${\bf z}(t+\tau)$ in tandem, resulting in the outputs $F(\mathbf{z}(t))$ and its time-lagged counterpart, $F(\mathbf{z}(t+\tau))$, which are then used for the calculation of the committor in \figref{fig:vcns_iterative} and the two-point correlation function in \eqref{eq:Cqq}. This architecture appears to form the basis of a reasonable strategy for data-driven learning of the committor~\cite{chen_chasing_2023,chen_discovering_2023}

In the course of the learning process, the correlation function $C_{qq}$ of \eqref{eq:Cqq} is minimized to obey the proposed variational principle. In order to impose continuity in the basins of the committor function, defined through scalar function $F$ via the relation in \figref{fig:vcns_iterative}, we need to minimize the following term,
\begin{equation}
    \mathcal{L}_b = \| F(\mathbf{z})\|^2_{\mathbf{z}\in A}+\| F(\mathbf{z})-1\|^2_{\mathbf{z}\in B},
\end{equation}
wherein $\| \cdot \|$ is the Euclidean norm. Hence, the variational principle of reference~\citen{chen_discovering_2023} is transformed to minimize the following total loss function,
\begin{equation}\label{eq:loss_q}
    \mathcal{L}_{q} = 2C_{qq} +k_{b}\mathcal{L}_b, 
\end{equation}
where $k_b$ is a positive constant to be adjusted upstream from the learning process.

%Moreover, a Siamese architecture~\cite{bromley_signature_1993,chicco_siamese_2021} is proposed for the evaluation of $F(\mathbf{z}(t))$ and its time-lag counterpart, $F(\mathbf{z}(t+\tau))$, since we need information of the time-evolution \chris{OF WHAT?} to recover the proper transitions encoded in the committor definition. 

\subsection*{The Committor-Consistent String}

The goal of the present work is to obtain the committor function and a string consistent with it, that will minimize the reaction flux with the final objective of carrying out enhanced-sampling simulations along a PCV, defined by such a CCS, $\{\widetilde{\bf z}_i\}_{i=1}^m$, the discretized form of which reads~\cite{branduardi_b_2007}
\begin{equation}\label{eq:def_s}
    s({\bf z}) = \sum_{k=1}^m {\frac{k-1}{m-1}}\sigma_k({\bf z}), \quad \sigma_k({\bf z})\equiv \frac{\displaystyle e^{\displaystyle -\alpha(\widetilde{\bf z}_k-{\bf z})^2}}{\displaystyle  \sum_{j}  e^{\displaystyle -\alpha(\widetilde{\bf z}_j-{\bf z})^2}},\quad \alpha\in\mathbb{R}^+.
\end{equation}
The CCS needs to connect both basins, so that $\widetilde{\bf z}_1\in A$ and $\widetilde{\bf z}_m\in B$ must be fixed, and it must follow the gradient of the committor, so that, it minimizes the reaction flux. 
For this purpose, and inspired by reference~\citen{he_committor-consistent_2022}, we will assume that the committor can be approximated by a given set of basis functions, $\{f_k\}_k$. However, instead of using the Voronoi tessellation of reference~\citen{he_committor-consistent_2022}, we have turned to a Gaussian expansion, in the spirit of PCVs\cite{branduardi_b_2007} (see \eqref{eq:def_s}), the functional form of which reads,
\begin{equation}\label{eq:def_expansion}
    q_s({\bf z})=\mathds{1}_A q_A + \mathds{1}_{A^c \cap B^c}s_q({\bf z}) + \mathds{1}_B q_B, \quad s_q({\bf z}) = \sum_{k=1}^m b_k \sigma_k({\bf z}), \quad  \bf{b} = \min_{\bf{b}^{\prime}} \| \bf{C}\bf{b}^{\prime}-\bf{q}\|,
\end{equation}
where $\mathds{1}_M$ is the indicator function of the set $M$,  $q_A=0$ and $q_B=1$, and ${\bf b}= (b_1,\dots,b_m)\in\mathbb{R}^{m}$, the set of coefficients computed to ensure the consistency with the committor  obtained by means of the VCN. Here, we impose that $q_i\equiv q(\widetilde{{\bf z}}_i)=s_q(\widetilde{{\bf z}}_i)$ for $i=1,\dots,m$, yielding a linear system of equations, $\bf{q} = \bf{C}\bf{b}$, with ${\bf q}=(q_1,\dots,q_m)\in\mathbb{R}^{m}$, and ${\bf C} = (\sigma_i(\widetilde{{\bf z}}_j))_{i=1,j=1}^{m}\in\mathcal{M}_{m\times m}$, a square, symmetric matrix. For $m$ sufficiently large, this system can be incompatible due to singularities in the matrix of coefficients, i.e., $\det{\bf{C}}=0$. Hence, to ensure a solution ${\bf b}$ for $\bf{q} = \bf{C}\bf{b}$, we propose to solve the system in a least-squares sense (see \eqref{eq:def_expansion}). It is worth noting that the expansion in terms of normalized Gaussians, \eqref{eq:def_expansion}, allows the committor to be bound in the $[0,1]$ interval. Consequently, the expansion map $q_s$ defined in \eqref{eq:def_expansion} is locally consistent with $q$. Furthermore, the tangent of the string aligns with the gradient of $q_s$. Specifically, at the midpoint between consecutive images, the gradient of $q_s$ is parallel to the string tangent when the constant $\alpha$ is large enough (see the SI for more details). In fact, whereas for the computation of the discrete form of the PCV (see \eqref{eq:def_s}) a value of $\alpha$ comparable with the inverse of the mean squared displacement between successive images is commonly utilized (see reference~\citen{branduardi_b_2007}), in the definition of the committor expansion (see reference~\eqref{eq:def_expansion}) a larger value ought to be used. A mathematically detailed discussion on selecting the value of $\alpha$ is provided in the SI.

The variational principle defined by the VCN should, therefore, also be obeyed by the string-like expansion defined in \eqref{eq:def_expansion}. Moreover, another optimization process is defined for the string learning based on a modified version of the original loss function,
\begin{equation}\label{eq:loss_s}
    \mathcal{L}_s = 2C_{q_sq_s}+k_b \mathcal{L}^{\prime}_b+k_{\mathrm{eq}}\mathcal{L}_{\mathrm{eq}}+k_{\mathrm{tr}}\mathcal{L}_{\mathrm{tr}},
\end{equation}
where $C_{q_sq_s}(\tau) = \langle (q_s(\tau)-q_s(0))^2\rangle/2$ and $\mathcal{L}^{\prime}_b = \| s_q({\bf z})\|_{{\bf z}\in A}+\| s_q({\bf z})-1\|_{{\bf z}\in B}$. In addition, $\mathcal{L}_{\mathrm{eq}}$ serves as a loss function that enforces equidistance between consecutive images (see SI). $\mathcal{L}_{\mathrm{tr}}$ coincides with $\sum_{i=1}^{m-1}(1-\cos\vartheta_i)/(m-1)$ when this value is large enough, where $\vartheta_i$ is the angle formed by the gradient of $q_s$ at $({\bf z}_{i+1}+{\bf z}_i)/2$ and the vector $ {\bf z}_{i+1}-{\bf z}_i$, is defined to ensure alignment of the gradient of $q_s$ with the tangent of the string (see SI). Finally, the constants $k_{\mathrm{eq}}$ and $k_{\mathrm{tr}}$ must be set prior to the learning process.

\subsection*{PCCANN and Iterative Learning Procedure}\label{subsec:iterative}

To identify the degrees of freedom that are important to understand the  biological process at play, we turn to an effective one-dimensional CV. This choice of such a seemingly simple reaction-coordinate model corresponds to a PCV defined by the CCS, and a CV perfectly apposite for enhanced-sampling simulations. To obtain a relevant CCS, we followed the iterative procedure depicted in \figref{fig:vcns_iterative}b to train the PCCANN. The iterative procedure starts from a rough initial guess of the transition pathway, for example, a set of images distributed on a straight line connecting the two basins. Under these premises, this initial string might be quite far from the relevant one. We then construct a second guess of the transition pathway, namely the CAP, from the trajectory produced in biased MD simulation using for the PCV the initial rough guess of the transition pathway. In order to build the CAP, the learned committor map is discretized into $N_\ell$ layers and bordered by consecutive isocommittor hyperplanes. Such hyperplanes are separated by the same width, $\Delta q$. Then, the CAP is constructed by averaging $\mathbf{z}$ of the trajectories in each layer of $q$. Next, a biased simulation is performed along the PCV defined by the CAP. Finally, the PCCANN is used to supply the committor and its corresponding CCS until convergence is reached. A schematic representation of this iterative process is provided in \figref{fig:vcns_iterative}b. 

It is worth noting that to this point, we have not yet considered the case where multiple, alternate transition pathways can be obtained. However, using CAPs, clusters of points populating distinct regions of CV space can be distinguished, as commented in \nameref{ssec:triple-well} and \nameref{ssec:chignolin}. Different clusters could imply different transition pathways and reactive tubes. Under these circumstances, a CCS will be sought for each possible transition pathway. In order to distinguish between alternative reactive tubes, we cluster the trajectory points, and compute a CAP for each subset of points. Next, each CAP corresponds to a distinct iterative procedure, as described in \figref{fig:vcns_iterative}b, from whence independently converged CCS are produced. 

Due to the use of biasing forces to enhance sampling in the iterative procedure, corrupting the underlying dynamics of the biomecular process at hand, we need to employ a reweighting technique\cite{wang_understanding_2020,rydzewski_multiscale_2021,chen_discovering_2023} to compute the unbiased time-correlation functions and all the corresponding estimates appearing in the loss functions (see \eqref{eq:loss_q} and \eqref{eq:loss_s}), following,
\begin{equation}\label{eq:rew}
    \langle f(\tau)g(0)\rangle = \frac{\displaystyle \langle e^{\delta W(\tau)/k_{\rm B}T}f(\tau)e^{\delta W(0)/k_{\rm B}T}g(0)\rangle_b}{\displaystyle e^{\delta W/k_{\rm B}T}},
\end{equation}
where subscript $b$ stands for the average using the biased ensemble, and $\delta W({\bf z})$ is the perturbation biasing potential. The expression in \eqref{eq:rew} provides an approximation valid only for small values of $\tau$, which were utilized (see SI) to generate the time-lagged trajectory in the Siamese ANN of the PCCANN.

%{\color{red} This paragraph should be in the final discussion and not in the method:} It is worth noting that to this point, we have not yet considered the case where multiple, alternate thransition pathways can be obtained. However, using CAPs, clusters of points populating distinct regions of CV space can be distinguished, as commented in \nameref{ssec:triple-well} and \nameref{ssec:chignolin}.
%Different clusters could imply different transition pathways and reactive tubes. Under these circumstances, a CCS will be sought for each possible transition pathway. In order to distinguish between alternative reactive tubes, we cluster the trajectory points, and compute a CAP for each subset of points. Next, each CAP corresponds to a distinct iterative procedure, as described in \figref{fig:vcns_iterative}b, from whence independently converged CCS are produced.

\bmhead{Codes and software}

The designed PCCANN has been built using the machine-learning library PyTorch\cite{paszke2019pytorch} based on the computational programming language Python\cite{python2009}. The trajectories are generated by hand-made programs for the simple cases of BS and MB, and the popular MD package, NAMD\cite{Phillips2020} and Colvars library\cite{Fiorin2013}, for the rest of the cases.

\bmhead{Data availability}

All the data in this work are provided upon request.

\bmhead{Code availability}

All the codes in this work are provided upon request.

%Conclusions may be used to restate your hypothesis or research question, restate your major findings, explain the relevance and the added value of your work, highlight any limitations of your study, describe future directions for research and recommendations. 

%In some disciplines use of Discussion or 'Conclusion' is interchangeable. It is not mandatory to use both. Please refer to Journal-level guidance for any specific requirements. 

\backmatter

\bmhead{Supplementary information}

The article has accompanying supplementary information.
%If your article has accompanying supplementary file/s please state so here. 

%Please refer to Journal-level guidance for any specific requirements.

\bmhead{Acknowledgments}

The European Research Council (project 101097272 ``MilliInMicro''), the Université de Lorraine through its Lorraine Université d'Excellence initiative, and the Région Grand-Est (project ``Respire'') are gratefully acknowledged for their support. The authors are indebted to Haochuan Chen for fruitful discussions and for his comments on the manuscript.



\section*{Supplementary material}

# Iterative variational learning of committor-consistent transition pathways using artificial neural networks

# Supplementary Information

Alberto Megías [1,2†], Sergio Contreras Arredondo [1†], Cheng Giuseppe Chen [1†], Chenyu Tang [1†], Benoît Roux [4], Christophe Chipot [1,3,4,5*]

[1*]Laboratoire International Associé Centre National de la Recherche Scientifique et University of Illinois at Urbana-Champaign, Unité Mixte de Recherche n°7019, Université de Lorraine, B.P. 70239, 54506 Vandœuvre-lès-Nancy cedex, France.
[2]Complex Systems Group and Department of Applied Mathematics, Universidad Politécnica de Madrid, Av. Juan de Herrera 6, E-28040 Madrid, Spain.
[3]Beckman Institute, University of Illinois at Urbana-Champaign, Urbana, USA.
[4]Department of Biochemistry and Molecular Biology, University of Illinois at Urbana-Champaign, Urbana, USA.
[5]Department of Physics, University of Chicago, Chicago, USA.

*Corresponding author(s). E-mail(s): chipot@illinois.edu;
†These authors contributed equally to this work.

# Theoretical Backdrop

## Proof of $\nabla q_s$ and the String Tangent Alignment

Let us assume the expansion for the committor in the string learning presented in equation (5) in the main text. Let be $\{\widetilde{\mathbf{z}}_k\}_{k=1}^{m}$ the string considered in the expansion and $\{\overline{\mathbf{z}}_k\}_{k=1}^{m-1}$ their middle points, i.e., $\overline{\mathbf{z}}_k = (\widetilde{\mathbf{z}}_k + \widetilde{\mathbf{z}}_{k+1})/2$. Hence,

$$\left.\frac{\partial}{\partial \mathbf{z}} q_s(\mathbf{z})\right|_{\mathbf{z}=\overline{\mathbf{z}}_k} = -2\alpha \sum_{j=1}^{m} b_j (\overline{\mathbf{z}}_k - \mathbf{z_j}) \sigma_j(\overline{\mathbf{z}}_k)(1 - \sigma_j(\overline{\mathbf{z}}_k)). \tag{1}$$

If $\alpha$ is sufficiently large, the dominating terms in the summation are those for $j = k, k+1$. Hence, taking into account that $\overline{\mathbf{z}}_k = \mathbf{z} + \Delta_k \mathbf{z}$ and $\overline{\mathbf{z}}_{k+1} = \mathbf{z} - \Delta_k \mathbf{z}$ and using $\sigma_k(\overline{\mathbf{z}}_k) = \sigma_{k+1}(\overline{\mathbf{z}}_k)$,

$$\left.\frac{\partial}{\partial \mathbf{z}} q_s(\mathbf{z})\right|_{\mathbf{z}=\overline{\mathbf{z}}_k} \approx 2\alpha \sigma_k(\overline{\mathbf{z}}_k)(1 - \sigma_k(\overline{\mathbf{z}}_k))(b_{k+1} - b_k) \Delta_k \mathbf{z} \propto \Delta_k \mathbf{z}, \tag{2}$$

where $\Delta_k \mathbf{z}$ is the discretized tangent of the string at $\overline{\mathbf{z}}_k$.

## Unbiasing Scheme

To illustrate the validity of the unbiasing scheme used in this work, we present the random-walk-like derivation presented in references 1–3. Assuming that the studied diffusive system, subjected to a biasing force, can be described by a jump-like process on a discrete-state Markov model, the change on the CVs after a time lag $\tau$ is reflected on the transition $\mathbf{z} \equiv \{z_1, \ldots, z_i, \ldots, z_m\} \to \mathbf{z}' \equiv \{z_1, \ldots, z_i + \Delta z, \ldots, z_m\}$. Hence, the transition probability of such a process would be

$$P_\tau(\mathbf{z}'|\mathbf{z}) = e^{-W(\mathbf{z}')/2k_\mathrm{B}T} \left(\tau \frac{D(\mathbf{z}') + D(\mathbf{z})}{2(\Delta z)^2}\right) e^{W(\mathbf{z})/2k_\mathrm{B}T}, \tag{3}$$

where $D(\mathbf{z})$ is the diffusion coefficient at a value $\mathbf{z}$ of the CVs and $W(\mathbf{z})$ is the potential of mean force (PMF). This approach coincides with a discretization of the Smoluchowski equation[4], valid for small $\Delta z$. Thus, if the PMF is perturbed by a position-dependent factor $\delta W$, the perturbed transition probability reads,

$$P_\tau^{(p)}(\mathbf{z}'|\mathbf{z}) = e^{-\delta W(\mathbf{z}')/2k_\mathrm{B}T} P_\tau(\mathbf{z}'|\mathbf{z}) e^{\delta W(\mathbf{z})/2k_\mathrm{B}T}, \tag{4}$$

which yields equation (7) in the main text. This approach is valid when the perturbation is small enough, that is, when $\tau$ is sufficiently small.

## String Method with Swarms of Trajectories

The string method[5] with swarms of trajectories[6,7] (SMwST) is a method used to compute a so-called "zero-drift pathway" (ZDP) for transitions between two states (A

and B) in complex, high-dimensional systems. The ZDP represents a path along which the system evolves on average without deviation in orthogonal directions, given that it is initially placed on this path and allowed to evolve freely for a short time. The SMwST finds the ZDP through an iterative process of the following steps, continuing until convergence:

1. **Initialization**. The pathway is discretized into $M$ images (or points) from state A to B in collective variable (CV) or conformational coordinate (CC) space. These initial images can be generated using steered molecular dynamics (SMD) or a direct linear interpolation between states in CV or CC space.
2. **Shooting swarm of trajectories**. For each image $i$ on the path, a swarm of NN short unbiased trajectories is generated over a time step $\delta\tau$, capturing local displacements under natural dynamics. The displacement of the $j$-th trajectory for image $i$, $\Delta z_{i,j}$ is recorded.
3. **Averaging displacement and updating the Path**. The average displacement $\langle \Delta z_i \rangle$ and standard deviation $\sigma(\Delta z_i)$ of each image $i$ are calculated. Each image $i$ is then updated based on the calculated displacement according to:

$$z'_i = z_i + \langle \Delta z_i \rangle + \alpha \sigma(\Delta z_i) X \qquad (5)$$

   where variable $X$ is randomly drawn from the standard normal distribution $\mathcal{N}(0,1)$, and $\alpha$ is a parameter scaling random fluctuation. $\alpha$ linearly decreases to zero to enable simulated annealing alike searching within the configurational space.[7]
4. **Re-parametrization**. The images along the path are re-parametrized to ensure equidistance.
5. **Equilibrium simulation**. After updating, each image is equilibrated by a simulation of time $\tau$ with its position restrained at $z'_i$. The equilibrated path then serves as the input for the next iteration, returning to Step 2.

Further details about the SMwST method can be found in references [6] and [7].

## Effect of the Choice of $\alpha$ in the Softmax Basis Set

To provide guidance for the choice of a suitable value of $\alpha$, let us follow the definition of the so-called *softmax* set of functions $\{\sigma_i\}_{i=1}^m$ (see equation (4) from the main text). This set of functions maps a point in the CV space into the (relative) interior of an $m-1$-dimensional unit simplex, i.e., $(0,1)^m$. The constant $\alpha$ controls the sharpness of the expansion, that is, in the limit $\alpha \to \infty$, the set of *softmax* functions converges to the *argmax* function point wise[8], which corresponds to the Voronoi tessellation proposed in reference [9]. We seek a continuous map, and, thus, need to work with a finite $\alpha$. However, if $\alpha$ is too small, the behavior of the expansion is inaccurate, i.e., in the limit $\alpha \to 0^+$, the *softmax* map converges point wise to the center of the simplex[8], which would prevent the images to evolve towards the production of a good committor. For this reason, we look for a large enough, albeit finite $\alpha$, so that the expansion scapes from the center of the simplex and goes towards a Voronoi-like tessellation, while preserving continuity. Hence, the typical use of the inverse of the mean squared displacement between successive images for the value of $\alpha$ is no longer

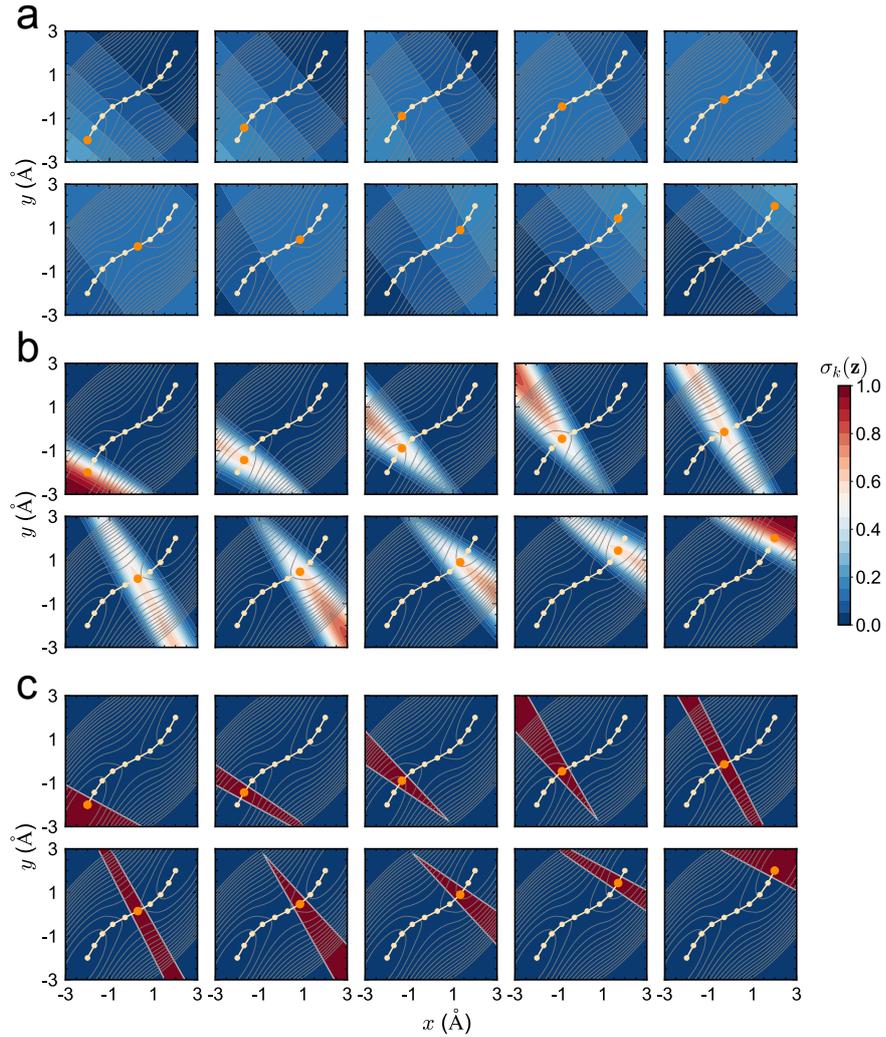

**Fig. 1**: Softmax functions $\{\sigma_k\}$ for each $k$ image of the CAP for the BS potential, using a value for $\alpha$ equal to $1/30$ (a), 1 (b) and 30 (c) times the inverse of the mean squared distance between consecutive images.

valid. To illustrate the former discussion, in Fig. 1, one can see a comparison between three different choices of the value of $\alpha$ using the committor-averaged path (CAP) of the unbiased isotropic Berezhkovskii–Szabo (BS) case with 10 images.

## Introduction of Geometric Restraints in the Loss Function

$\mathscr{L}_{\text{eq}}$ **Restraint.** Our aim during the learning of the CCS is to obtain a string, the images of which are as equidistant as possible, but such that they have some leeway

to evolve during this process. We, therefore, propose the following addition to the loss function,

$$\mathscr{L}_{\text{eq}} = \begin{cases} \sqrt{\text{Var}[\Delta_k \widetilde{\mathbf{z}}]}, & \text{if } \sqrt{\text{Var}[\Delta_k \widetilde{\mathbf{z}}]} \geq C_{\text{eq}} \text{E}[\Delta_k \widetilde{\mathbf{z}}], \\ 0, & \text{otherwise,} \end{cases} \quad (6)$$

where $\Delta \widetilde{\mathbf{z}} = \|\widetilde{\mathbf{z}}_k - \widetilde{\mathbf{z}}_{k+1}\|$, with $\widetilde{\mathbf{z}}_k$ being the $k$-th image of the string being learned, Var$[\cdot]$ standing for the variance and E$[\cdot]$ for the mean, and $C_{\text{eq}} > 0$ being a constant that controls the freedom given to the images to depart from being exactly equidistant.

**$\mathscr{L}_{\text{tr}}$ Restraint.** Since the alignment between the gradient of the committor expansion and the tangent to the string is formally fulfilled in the limit of large $\alpha$ values, imposing this property is desirable when seeking the CCS. Hence, we defined the contribution of the loss function that ensures this requirement as $\sum_{i=1}^{m-1}(1 - \cos \vartheta_i)/(m-1)$, where $\vartheta_i$ is the angle formed by the gradient of $q_s$ at $(\mathbf{z}_{i+1} + \mathbf{z}_i)/2$ and the vector $\mathbf{z}_{i+1} - \mathbf{z}_i$. However, in the spirit of the previous section, in order to grant the string some leeway for excursions, this contribution is in practice defined as,

$$\mathscr{L}_{\text{tr}} = \begin{cases} \sum_{i=1}^{m-1}(1 - \cos \vartheta_i)/(m-1), & \text{if } \sum_{i=1}^{m-1}(1 - \cos \vartheta_i)/(m-1) \geq C_{\text{tr}}, \\ 0, & \text{otherwise.} \end{cases} \quad (7)$$

## Computational Details

### PCCANN Architecture

The PCCANN architecture consists of a artificial neural network (ANN) combined with an external layer used to optimize the images of the string. The ANN is constructed with an input layer containing a number of neurons equal to that of the input features, followed of four dense layers with 32 neurons each. ELU was used as the activation function for the hidden layers. The ANN is set with an output layer of one neuron to produce the final output, $F(\mathbf{z})$. The external layer is defined with the same number of neurons as the images in the string. For training, the initial trajectory is split into a training set and a validation set in a 9:1 ratio. Both, the variational committor network (VCN) and the nodes associated with the string are trained using the Adam optimizer[10] over a maximum of around 3000-5000 epochs, during which the training set is randomly divided into batches of size $\lfloor N_{\text{train}}^{0.6} \rfloor$, where $N_{\text{train}}$ is the number of samples in the training set. Optimization is stopped if the validation loss does not decrease for 100 consecutive epochs, applying an early stopping strategy. To ensure effective learning of the string-image positions, a preliminary training of 50 epochs is done for the committor map using the VCN part of the ANN. Throughout the training, the value of $\alpha$ is set to be higher than the inverse of the mean squared distance between successive images of the string in the previous iteration. This choice of $\alpha$ ensures that the basis set effectively covers the space, leading to accurate learning of the committor map.

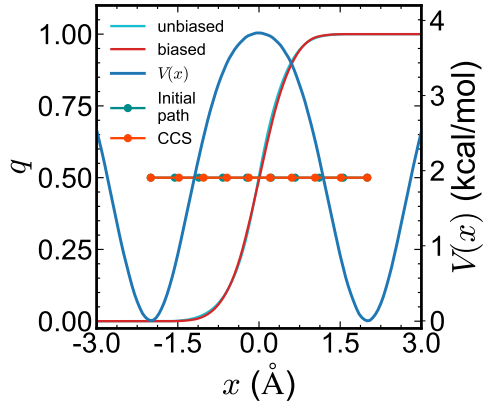

**Fig. 2**: Committor function, $q$, learned by the PCCANN using a biased simulation (red), and by the VCN using an unbiased simulation (light blue), along with the initial guess path (green) and the CCS (orange). The double-well potential energy profile, $V(x)$, is shown in blue.

## One-Dimensional Double-Well Potential

The one-dimensional double-well (DW) potential is here defined as:

$$V(x) = -4e^{-(x+2)^2} - 4e^{-(x+2)^2} + 4 \qquad (8)$$

To learn the committor and the CCS, an unbiased and a WTM-eABF simulation of 125 ns each were carried out with a timestep of 0.5 fs. The first 25 ns of the biased trajectory were discarded in order to guarantee the convergence of the perturbation biasing potential, $\delta W(\mathbf{z})$[11]. The temperature was kept at 300 K through the use of Langevin dynamics.

The PCCANN parameters were set to $k_b = 1$, $k_{\text{eq}} = 1$, $k_{\text{tr}} = 1$, $m = 10$ images, $\alpha = 3/\text{MSD}(\{\widetilde{\mathbf{z}}_\mathbf{k}\}_k)$, $C_{\text{eq}} = 0.1$ and $C_{\text{tr}} = 0.05$. The maximum number of epochs was set to 2000 with an early stropping set at 100 epochs.

This system served as a first benchmark to test the ability of the model to learn the committor function without the added difficulty of simultaneously learning the path. Our results show that $q(\boldsymbol{x})$ learned by our model using a biased simulation is virtually identical to our reference, $q(\boldsymbol{x})$ obtained using the VCN in conjunction with an unbiased trajectory (see Fig. 2).

## Berezhkovskii-Szabo Two-Dimensional Potential

To illustrate the effectiveness of the committor-consistent path finding iterative algorithm, we try to align a path linearly interpolated between the two metastable states of the Berezhkovskii-Szabo two-dimensional potential[12] with anisotropic diffusivities to the gradient of $q(\mathbf{z})$. The potential energy function is

$$\beta V(x,y) = \beta V(x) + 1.01\omega^2(x-y)^2/2 \qquad (9)$$

where $\beta$ is $1/k_B T$ and $V(x)$ is defined as

$$\beta V(x) = \begin{cases} -\omega^2 x_0^2/4 + \omega^2(x+x_0)^2/2, & x < -x_0/2 \\ -\omega^2 x^2/2, & -x_0/2 \leq x \leq x_0/2 \\ -\omega^2 x_0^2/4 + \omega^2(x-x_0)^2/2, & x_0/2 < x \end{cases} \quad (10)$$

where $\omega^2$ is 4, and $x_0$ is 2.2 Å, respectively. In each iteration, 250 ns trajectories of $x$ and $y$ for learning the committor were drawn from an overdamped Langevin dynamics simulation of a carbon atom evolving on $V(x,y)$ considering isotropic ($\gamma_x/\gamma_y = 1$) and anisotropic diffusion for $\gamma_x/\gamma_y = 0.1$ and $\gamma_x/\gamma_y = 10.0$, where $x$ and $y$ are the atomic coordinates along the abscissa and the ordinate, respectively. The biased trajectories used to feed the NNs models where obtained from WTM-eABF simulations of a carbon atom evolving on $V(x,y)$, where $x$ and $y$ are the atomic coordinates, where the first 25 ns are discarded for the learning. We defined all the PCVs for each iteration to be defined by 12 equidistant images. The end-points of all the strings defining the different PCVs used for the biased simulations are fixed to connect the two metastable states at $(x,y) = (-2,-2)$ and $(x,y) = (2,2)$.

The values of the different constants in the loss function defined in the PCCANN were $k_b = 100$, $k_{\text{eq}} = 1$, $k_{\text{tr}} = 1$, $m = 30$ images, $\alpha = 30/\text{MSD}(\{\widetilde{\mathbf{z}_k}\}_k)$, where $\text{MSD}(\{\widetilde{\mathbf{z}_k}\}_k)$ is the mean squared displacement of the guess string, and $C_{\text{eq}} = 0.1$ and $C_{\text{tr}} = 0.3$ (see their definitions in equation (6) and equation (7), respectively). In addition, the maximum number of epochs was set to 5000 with an early stropping of 100 epochs.

## Müller-Brown Two-Dimensional Potential

The two-dimensional MB potential[13] is particularly challenging for optimization algorithms due to its complex landscape. The MB potential energy, $V(x,y)$, is defined as,

$$V(x,y) = k \sum_{i=1}^{4} d_i e^{a_i(x-x_i)^2 + b_i(x-x_i)(y-y_i) + c_i(y-y_i)} \quad (11)$$

where k=0.05, $[d_1, d_2, d_3, d_4] = [-200, -100, -170, 15]$, $[a_1, a_2, a_3, a_4] = [-1, -1, -6.5, 0.7]$, $[b_1, b_2, b_3, b_4] = [0, 0, 11, 0.6]$, $[c_1, c_2, c_3, c_4] = [-10, -10, -6.5, 0.7]$, $[x_1, x_2, x_3, x_4] = [1, 0, -0.5, -1]$ and $[y_1, y_2, y_3, y_4] = [0, 0.5, 1.5, 1]$. $x$ and $y$ are the atomic coordinates along the abscissa and the ordinate, respectively. The units of $x$, $y$ and $V(x,y)$ are Å, Å, and kcal/mol, respectively.

We are using, as in previous cases, $x$ and $y$ as the input for the NN models. The two metastable states are defined as the point sets $\{(x,y) \,|\, x \in (-1.0, -0.2), y \in (1.0, 2.0), V(x,y) < 1.0\}$ and $\{(x,y) \,|\, x \in (0.2, 1.0), y \in (-0.2, 2.0), V(x,y) < 3.0\}$, respectively. This system is commonly use to assess the ability of algorithms to navigate complicated energy surfaces, identify critical points, and trace reaction pathways. Its topology, provides an ideal platform for testing the robustness and efficiency of various methods aimed at solving problems in MD and chemical reactions.

To highlight one of the biggest challenges in working with complex systems, where the surface on which the dynamics happen is usually unknown, we mapped the two-dimensional MB system onto a Swiss roll structure[14] to have a new 3D coordinate system. We transformed the simulation trajectories using the Swiss roll mapping;

$$\begin{bmatrix} x \\ y \end{bmatrix} \rightarrow \begin{bmatrix} (c+x)\cos((c+x)f) \\ y \\ (c+x)\sin((c+x)f) \end{bmatrix} \quad (12)$$

where $f$ is the frequency and reflects how tightly the roll is wound and $c$ is a parameter to control range of $x$. In this work we are using $f = 3.0$ and $c = 1.8$, to ensure the correct range of $x$ and an enough challenging orientation and tightness of the Swiss roll.

Convergence of the learning process was achieved with the following values of the PCCANN parameters: $k_b = 1$, $k_{eq} = 1$, $k_{tr} = 1$, $m = 30$ images, $\alpha = 3/\text{MSD}(\{\widetilde{\mathbf{z}_k}\}_k)$, where $\text{MSD}(\{\widetilde{\mathbf{z}_k}\}_k)$ is the mean squared displacement of the guess string, and $C_{eq} = 0.1$ and $C_{tr} = 0.16$. In addition, the maximum number of epochs was set to 3000 with an early stropping of 100 epochs.

## Triple-Well Potential

The triple-well potential is a function commonly used to describe systems with multiple equilibrium states, each separated by energy barriers that define rare transitions[15,16]. In the context of physical systems and stochastic models, this potential is crucial for understanding metastability, where particles or systems remain in low-energy states for extended periods before transitioning to other states. This potential is described by the following equation,

$$\begin{aligned} V(x,y) = & \; 3e^{-x^2-\left(y-\frac{1}{3}\right)^2} - 3e^{-x^2-\left(y-\frac{5}{3}\right)^2} \\ & - 5e^{-(x-1)^2-y^2} - 5e^{-(x+1)^2-y^2} \\ & + \frac{2}{10}x^4 + \frac{2}{10}\left(y-\frac{1}{3}\right)^4 \end{aligned} \quad (13)$$

The triple-well potential, as described by equation (13), contains three wells: two deep minima at approximately $(\pm 1, 0)$ and a shallower minimum near $(0, 1.5)$. Additionally, the potential features three saddle points at $(\pm 0.6, 1.1)$, $(0.0, -0.4)$, and a maximum at $(0.0, 0.5)$. This structure represents a complex energy landscape, providing multiple possible paths for transition between states.

The PCCANN parameters for the upper channel were set to $k_b = 1$, $k_{eq} = 1$, $k_{tr} = 1$, $m = 30$ images, $\alpha = 3/\text{MSD}(\{\widetilde{\mathbf{z}_k}\}_k)$, where $\text{MSD}(\{\widetilde{\mathbf{z}_k}\}_k)$ is the mean squared displacement of the guess string, and $C_{eq} = 0.1$ and $C_{tr} = 0.7$. In contrast, for the lower channel these values are $k_b = 1$, $k_{eq} = 1$, $k_{tr} = 1$, $m = 30$ images, $\alpha = 3/\text{MSD}(\{\widetilde{\mathbf{z}_k}\}_k)$, $C_{eq} = 0.1$ and $C_{tr} = 0.1$. In addition, the maximum number of epochs was set to 5000 with an early stopping of 100 epochs.

As can be seen in Fig. 3, two distinct reactive tubes are identified at $T = 300\,\text{K}$ employing K-means clustering[17]. Owing to the mathematical form of the TW potential, the resulting paths are quite similar to the CAPs obtained from a naive partitioning of the CV space by a straight line at $y = 0$ Å, as commented on in the main text.

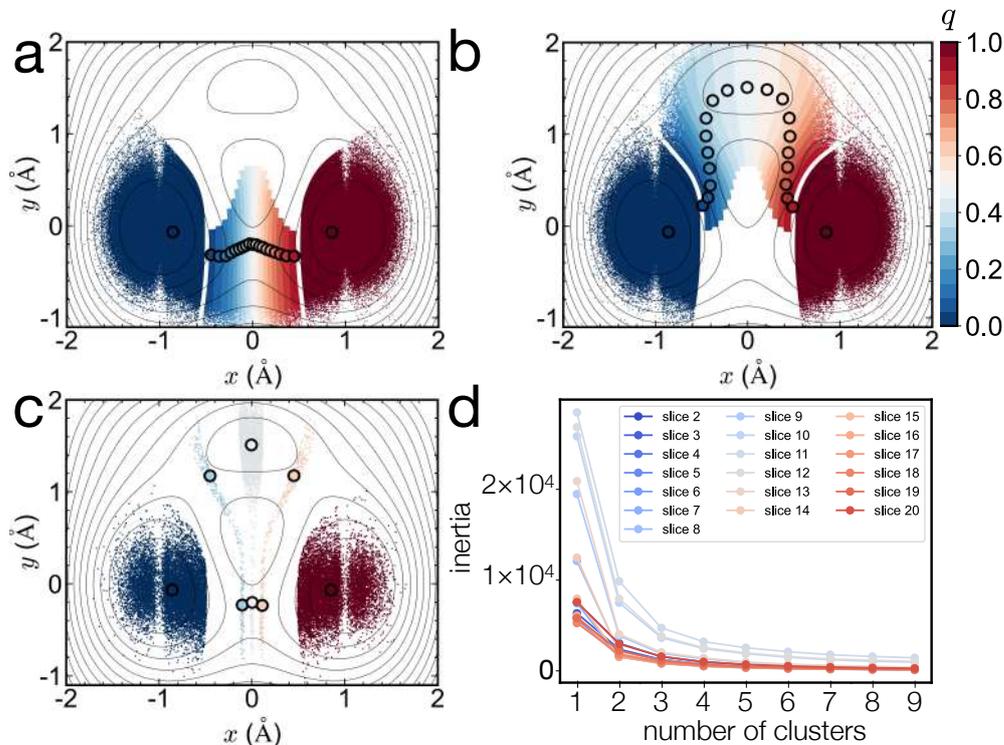

**Fig. 3**: Partitioning of the CV space sampled by a biased MD simulation at 300 K along a straight path connecting the two lower basins of the TW potential. The pathways highlighted as black circles is formed by the centroids of the points lying in finite slices of the committor map discretized for the lower (a) and upper (b) reactive tubes identified by K-means clustering[17]. Focus on three isocommittor slices to illustrate K-means clustering (showing only every 50 points sampled in the simulation to visually highlight the identified clusters) (c). Inertia computed with different numbers of clusters for each isocommittor slice of the committor map (d) The first ($q = 0$) and last ($q = 1$) isocommittor slices were excluded from the clustering analysis, since the same end-points are imposed for the two pathways. By employing the elbow method, presence of two reaction tubes was established.

9

### $N$–Acetyl–$N'$–methylalanylamide

For NANMA, the data for each iteration was provided by 125-ns WTM-eABF PCV simulations, of which the first 25 ns were discarded to guarantee the convergence of the perturbation biasing potential, $\delta W(\mathbf{z})$[11]. The system was described using the CHARMM22 force field and the temperature was kept at 300 K using the Langevin thermostat.

For both the two and four-dimensional cases, the PCCANN parameters were set to $k_b = 1$, $k_{eq} = 1$, $k_{tr} = 1$, $m = 15$ images, $C_{eq} = 0.2$ and $C_{tr} = 0.05$. For the two-dimensional case, $\alpha = 3/\text{MSD}(\{\widetilde{\mathbf{z}}_\mathbf{k}\}_k)$, while for the four-dimensional case, $\alpha$ was set equal to $30/\text{MSD}(\{\widetilde{\mathbf{z}}_\mathbf{k}\}_k)$. For the first iteration only (i.e., starting from the CAP to learn the fist CCS), in the two-dimensional case, a value of $C_{tr} = 0.12$ was used to allow the string to evolve more freely. Moreover, the maximum number of epochs was set to 2000 with an early stropping set at 100 epochs.

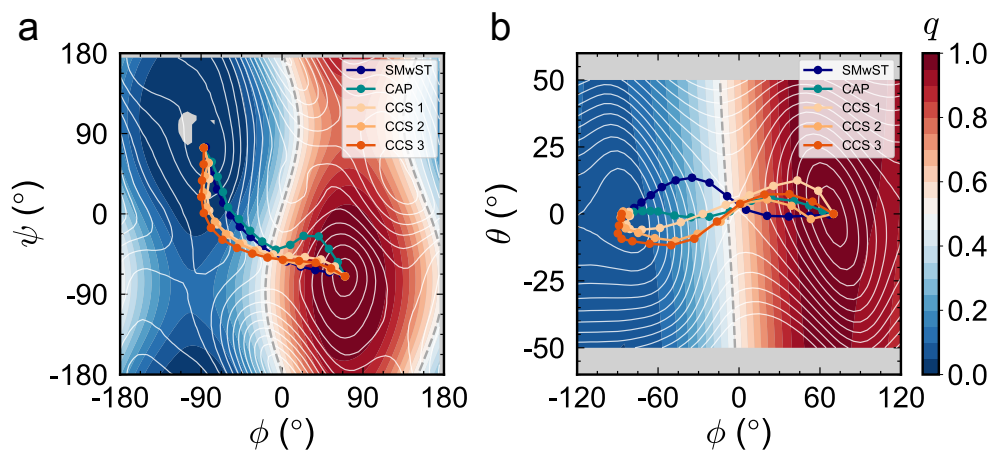

**Fig. 4**: Learned committor $q$ and iterated CCSs for NANMA using $(\phi, \psi, \theta, \omega)$ as the input features. The following variables have been kept constant when calculating the committor: $\theta = 2°$, $\omega = 0°$ (a); $\psi = 50°$, $\omega = 0°$ (b).

**Modified $\mathscr{L}_{tr}^*$ Restraint.** While the VCN, and, therefore, the PCCANN, is able to learn a reasonable committor function, the derivative of the latter is not always accurate. This is evident by observing the tendency of the VCN to learn values of $F(\mathbf{z})$ which are mildly outside of the $[0, 1]$ range. Inaccuracies in $\nabla F((z))$ are, however, not limited to these outlier regions (see Fig. 6e-f). Such artifacts, while minor in terms of modulus, can carry a dramatic effect on the CCS. Since $L_{tr}$ forces the learned path to follow the local $\nabla F((z))$, regardless of its magnitude, due to the stochastic nature of the optimization process, learning can easily get stuck in a local minimum of the loss function. For simple systems, i.e., either with nonperiodic CVs, or with a lower dimensionality, these noisy regions mainly lie in the A and B basins and regions of the conformational space that are not sampled, and ultimately do not overlap with

substantial portions of the learned path (see Fig. 6a-d). As can be seen, in the case of NANMA, this assumption, however, does not hold true.

When only $(\phi, \psi)$ are considered, this issue can be assuaged by running multiple independent trainings with the same trajectory, and only choosing the model which provides the lowest value of $\mathscr{L}_s$, i.e., according to the variational principle. Conversely, when $\theta$ and $\omega$ are also included, this strategy can easily break down, leading to a considerable increase in the number of trainings required. For this reason, a modified restraint was used, based on the general ideas behind $\mathscr{L}_{tr}$:

$$\mathscr{L}_{tr}^* = \sum_{i=1}^{m-1}(1-\cos\vartheta_i)\frac{w_j}{\sum_{j=1}^{m-1}w_j}, \quad w_i = \sqrt{\|\nabla F((\widetilde{\mathbf{z}}_{i+1}+\widetilde{\mathbf{z}}_i)/2)\|} \qquad (14)$$

Using equation (14), points with low modulus of $\nabla F$ (which is the case for the region where the artifacts are present) will contribute less to the loss function. At the same time, we decided to use the square root of the modulus to weight the contributions because of the wide range of possible values of $\nabla F$ (see Fig. 6). As a result, points with low $\nabla F$ will not be completely neglected in the learning process, striking what we found out to be a good balance.

## Chignolin

Chignolin immersed in a box of 2748 TIP3P water molecules was described using the Amber ff14SB force field[18]. Two Na$^+$ ions were added to ensure electric neutrality. Hydrogen mass repartitioning was applied in order to increase the integration time step to 4 fs. The temperature and pressure were kept constant at 340 K and 1 atm, respectively, through the use of Langevin dynamics and the Langevin piston algorithm[19].

For the straight path, a total of 2.6 $\mu$s of sampling was produced. For the first path, we gathered 2.5 $\mu$s for the CAP and 0.9 $\mu$s for the first CCS. As for the second

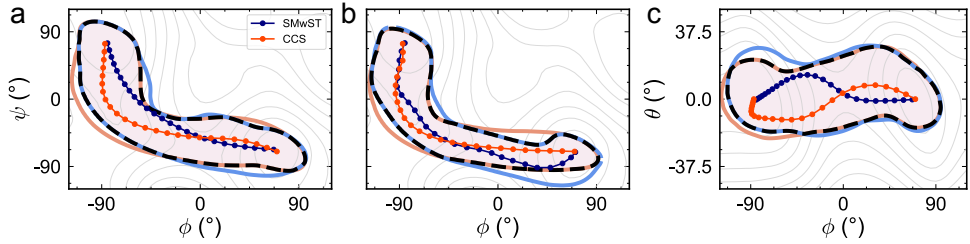

**Fig. 5**: Sampled region from biased PCV simulations using the CCS and the SMwST string obtained considering as the CVs $(\phi, \psi)$ (a) and $(\phi, \psi, \theta, \omega)$ (b,c) ($\theta = 2°$, $\omega = 0°$ (b); $\psi = 50°$, $\omega = 0°$ (c)). The orange and blue contours represent regions encompassing 95% of the total samples in the simulations following the CCS and the path supplied by the SMwST, respectively, whereas the black dashed contour indicates the overlapping sampling region.

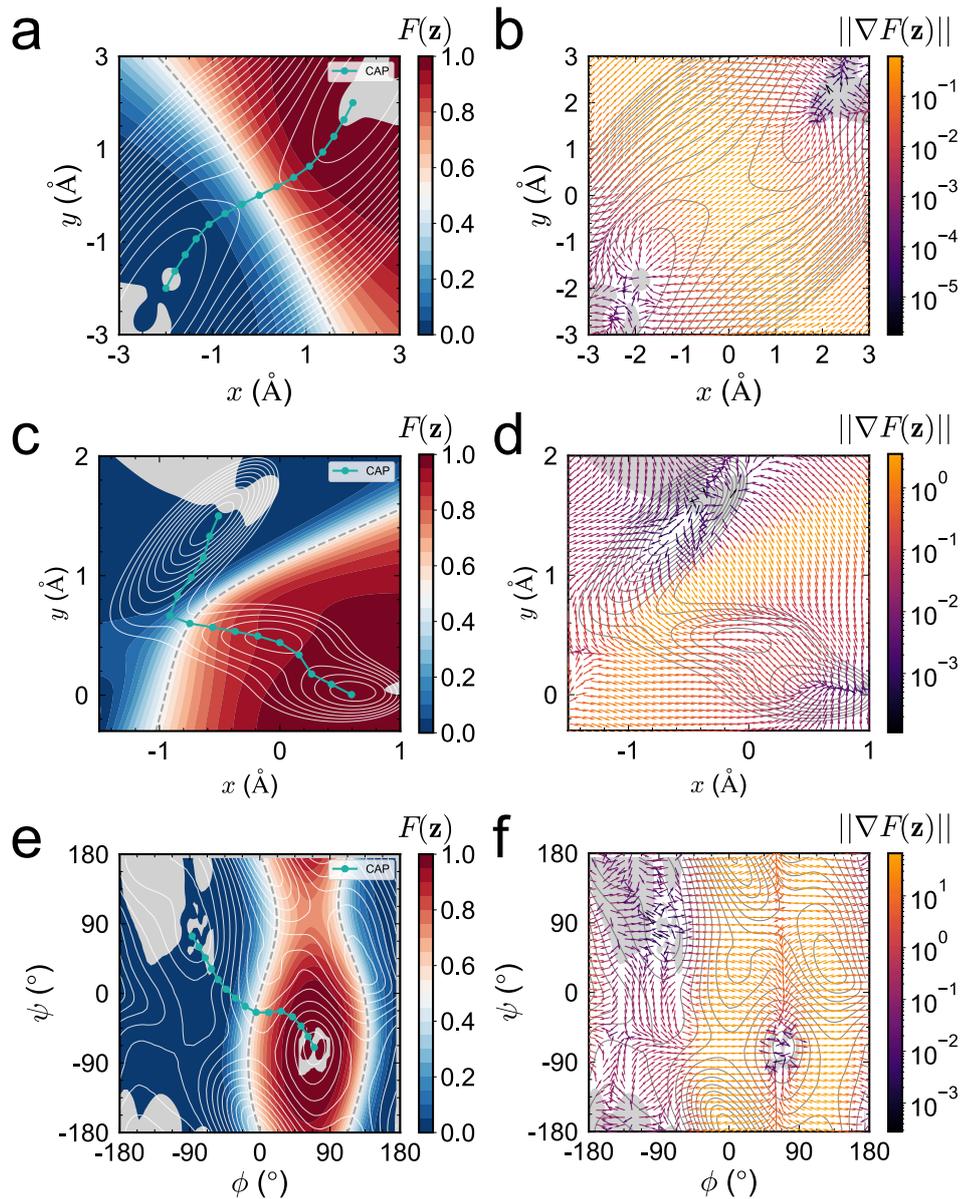

**Fig. 6**: Learned $F(\mathbf{z})$ and $||\nabla F(\mathbf{z})||$ from the VCN from an unbiased trajectory of BS ($\gamma_x/\gamma_y = 1.0$) (a,b), MB (c,d) and biased trajectory of NANMA sampling the ($\phi, \psi$) subspace (e,f). The CAPs used for the iterative learning are also shown to give an idea of what regions of the conformational spaces is relevant in the learning process (i.e., the regions where having an accurate estimation of $||\nabla F(\mathbf{z})||$ is critical).

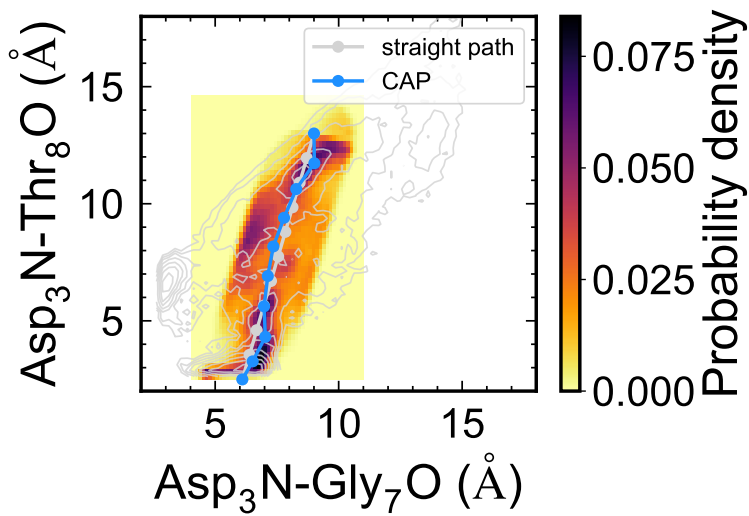

**Fig. 7**: Probability density distribution of the conformations sampled in the PCV simulation using the straight path connecting the folded and unfolded states.

path, we sampled over 3.4 $\mu$s for the CAP, 1.8 $\mu$s for the first CCS and 4.4 $\mu$s for the second CCS. Of these trajectories, only the last 400-800 ns were used to train the model, based on the convergence of $\delta W(\mathbf{z})$.

For both pathways, the PCCANN parameters were set to $k_b = 1$, $k_{eq} = 1$, $k_{tr} = 1$, $m = 10$ images, $\alpha = 3/\text{MSD}(\{\widetilde{\mathbf{z}}_\mathbf{k}\}_k)$. For the first iteration only, $C_{eq} = 0.4$ and $C_{tr} = 0.12$, while for all the other iterations they were set as 0.1 and 0.05, respectively. The maximum number of epochs was set to 5000 with an early stropping set at 100 epochs. The string provided as the starting point for the learning process does not coincide exactly with the path used for the PCV simulation. The latter was, in fact, extended slightly beyond the end-points of the reaction-coordinate model in order to prevent the artifacts occurring at the boundaries of the PCV, likely to affect the sampling of the folded and unfolded basins.

**Bifurcation in the Transition Path.** The probability density distribution of the conformations sampled during the PCV simulation along the straight path reveals multiple reactive tubes, which is further emphasized when calculating the CAP using the whole trajectory. This CAP remains very close to the straight path, and does not reflect the sampling distribution (see Fig. 7). Since different clustering algorithms yield different answers, it was chosen to split naively the individual configurations of the trajectory on the left or on the right side of the straight path in order to obtain the two CAPs that actually reflect the sampling distribution (see the main text).

**Free-Energy Calculations** The PMFs calculated for the two alternative paths are shown in Fig. 8. While the results for the first path are corroborated by previous

works using different approaches[20–22], the second path does not completely match the first one in terms of the free-energy difference between the folded and unfolded states, as well as the barrier separating them, somewhat lower than excepted, when considering the two-dimensional free-energy landscape projected onto $Asp_3N$-$Gly_7O$, $Asp_3N$-$Thr_8O$ [20,22].

One possible reason for this discrepancy may lie in the free-energy method utilized. It has been shown that WTM-eABF (and other enhanced-sampling algorithm, like replica-exchange umbrella sampling, or REUS, used in reference [20]) struggle to reach convergence when applied to chignolin, even after 2 $\mu$s of sampling[22]. Gaussian-accelerated WTM-eABF (GaWTM-eABF) appears to generate a more homogeneous sampling[22], but, for the sake of consistency with the other systems examined in this work, it was not employed.

Another source of discrepancy may be the PCV simulation itself. In order to explore the two paths independently, a strong restraint on the ancillary PCV $\zeta$, orthogonal to $s(\mathbf{z})$, was imposed. The objective of sampling the reactive tubes independently was not entirely attained, considering how in the simulation of the second path, a lot of time seems to be spent in trying to access the 1st path denied by the restraint (see the higher density of points at the point of the bifurcation in Fig. 9). The first path, in fact, seems to be the more probable one, since it is also more sampled than the second path in the PCV simulation along the straight path, see Fig. 7.

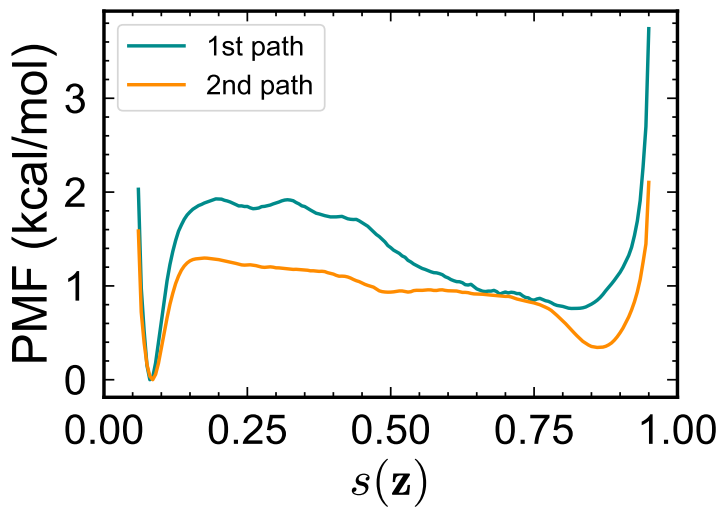

**Fig. 8**: PMFs of chignolin along the CCSs corresponding to the two transition paths.

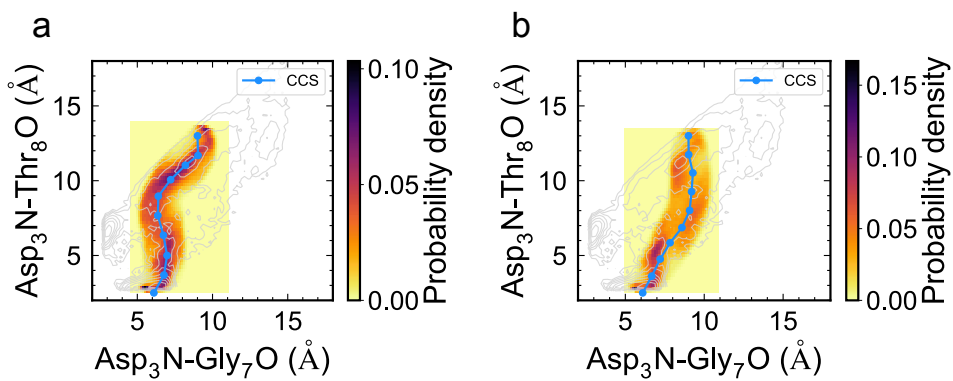

**Fig. 9**: Probability density distribution of the conformations sampled in the PCV simulation using the converged CCSs for the first (a) and the second (b) reaction tube.

**Clustering of the Transition-State Conformations.** K-Medoids clustering analysis[23] was performed to obtain conformations representative of the transition states. We labeled the sampled conformations as transition state when the committor predicted by the PCCANN was in the range of $[0.485, 0.525]$. The distances between the heteroatoms of $Asp_3$ and $Thr_6$, $Gly_7$ and $Thr_8$ (see Fig. 10) were computed and used for clustering purposes. The optimal number of clusters (four for both paths) was chosen by calculating the inertia as a function of the number of clusters and by employing the so-called elbow method (see Fig. 11).

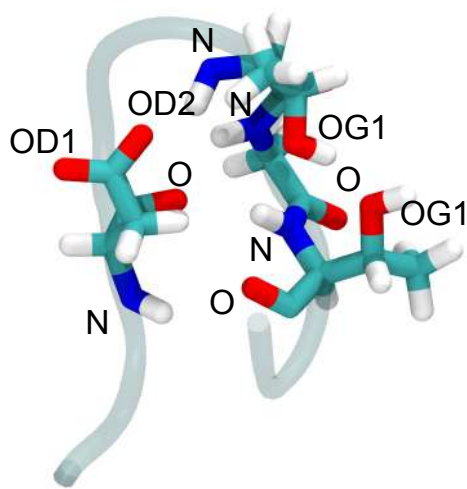

**Fig. 10**: PDB names of the heteroatoms considered in the clustering analysis.

15

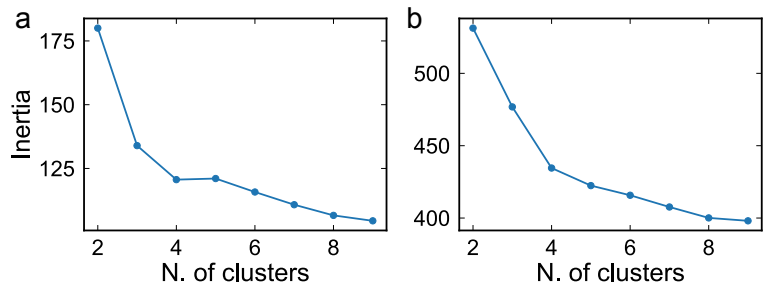

**Fig. 11**: Calculated inertia from the models obtained using K-medoids clustering the conformations belonging to the transition state of the first (a) and second (b) path.

Analyzing the distances between the heteroatoms of $Asp_3$ and $Thr_6$, $Gly_7$ and $Thr_8$ (see Fig. 12), we can broadly cluster the sampled transition-state conformations further in groups wherein $Asp_3$ mainly interacts with $Thr_6$ (see Fig. 12d,e,h), and groups wherein additional interactions between $Asp_3$ and $Thr_8$ are formed (see Fig. 12b,c,f,g). This classification generally agrees with the results from a recent computational study[24]. Moreover, it is worth noting that one of the clusters from the first path does not feature long-lived interactions between residues common to the conformations belonging to it. This outlier may stem from an erroneous inclusion of conformations of the transition-state ensemble, possibly as a result of inaccuracies in the learned committor function.

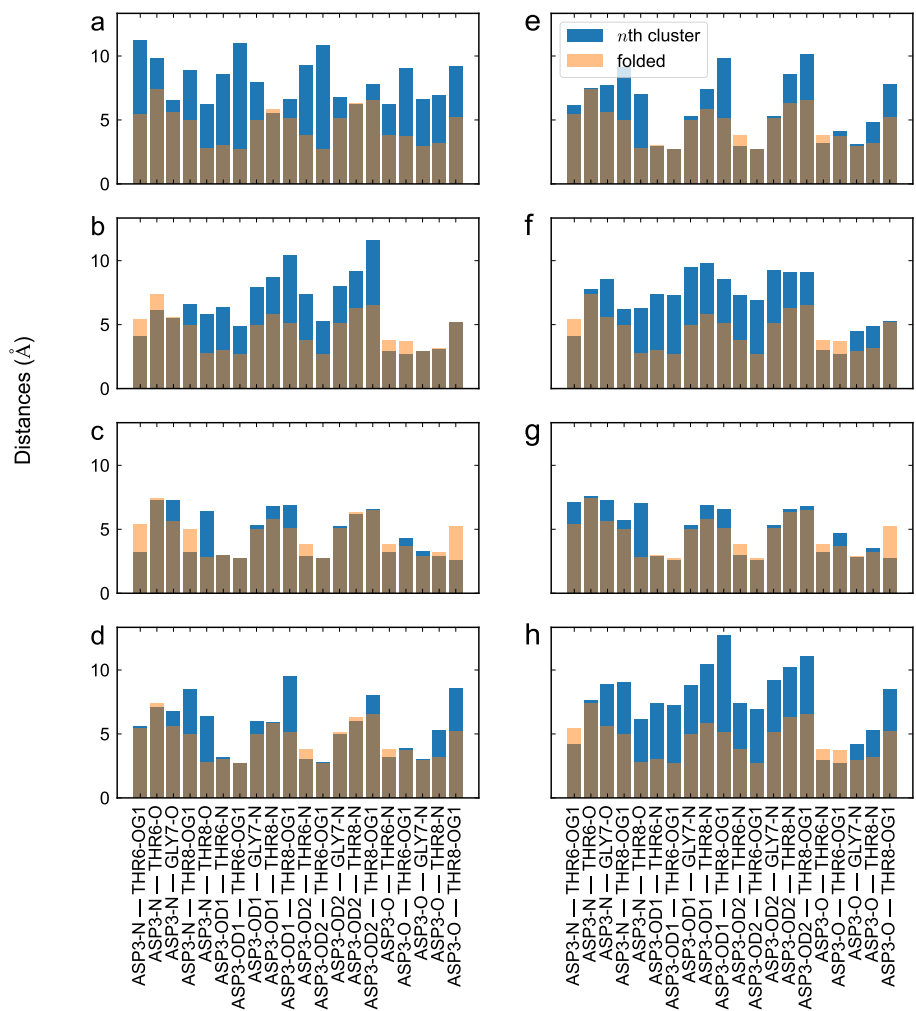

**Fig. 12**: Most frequent distances connecting the atoms involved in possible hydrogen bonds between $Asp_3$ and $Thr_6$, $Gly_7$ and $Thr_8$ in the conformations belonging to the different clusters of the transition state in the first (a-d) and second path (e-h), depicted in blue. The corresponding values calculated for the folded state are shown in orange in each panel.

## Transition Rates

### Transition Rates from PCCANN

For all the illustrations reported here, we assume a two-state Markov jump between metastable states $A$ and $B$,

$$A \underset{k_{BA}}{\overset{k_{AB}}{\rightleftharpoons}} B \qquad (15)$$

from which a transition rate can be determined. The equilibrium state probabilities are $p_A = k_{BA}/(k_{AB} + k_{BA})$ and $p_B = k_{AB}/(k_{AB} + k_{BA})$. The unidirectional fluxes are equal at equilibrium, i.e., $J_{AB} = p_A k_{AB} = J_{BA} = p_B k_{BA}$. The mean first passage time (MFPT) from $A$ to $B$ can be determined as $1/k_{AB}$.[25–27] From the PCCANN, in the spirit of the VCN[1], the transition rate can be obtained from $A$ to $B$ from the unidirectional flux $k_{AB} = J_{AB}/p_A$, where the flux, $J_{AB}$, is given by:

$$J_{AB} = \frac{1}{2\tau} \left\langle (q(\tau) - q(0))^2 \right\rangle = \frac{C_{qq}(\tau)}{\tau} \qquad (16)$$

Here, we determine the transition rate, $k_{AB}$, from the slope of the time-correlation function $C_{qq}(\tau)$, as a function of $\tau$ and $p_A$, for the different systems examined.

### Transition Rates from Position-Dependent Diffusivity

For comparison, we employed an alternate approach to calculate the transition rate, $k_{AB}$, from the potential of mean force (PMF) along the string, $\Delta G(s(\mathbf{z}))$, and the position-dependent diffusivity $D(s(\mathbf{z}))$, initially proposed by Pontryagin et al[28–30]:

$$k_{AB} = \left\{ \int_{s_A}^{s_B} ds \frac{e^{\beta \Delta G(s)}}{D(s)} \times \int_{s_A}^{s^*} ds e^{-\beta \Delta G(s)} \right\}^{-1} \qquad (17)$$

where $s^*$ is selected to be between $s_A$ and the nearest free-energy barrier. $\Delta G(s(\mathbf{z}))$ is obtained by sampling along PCVs[31] and $D(s(\mathbf{z}))$ is determined from Bayesian inferences[32] using the DiffusionFusion code[33]. The transition rates from the position-dependent diffusivity are shown in Table 1 for the two-dimensional Müller-Brown (MB), triple-well (TW) potentials, and NANMA.

## One-Dimensional Double-Well Potential

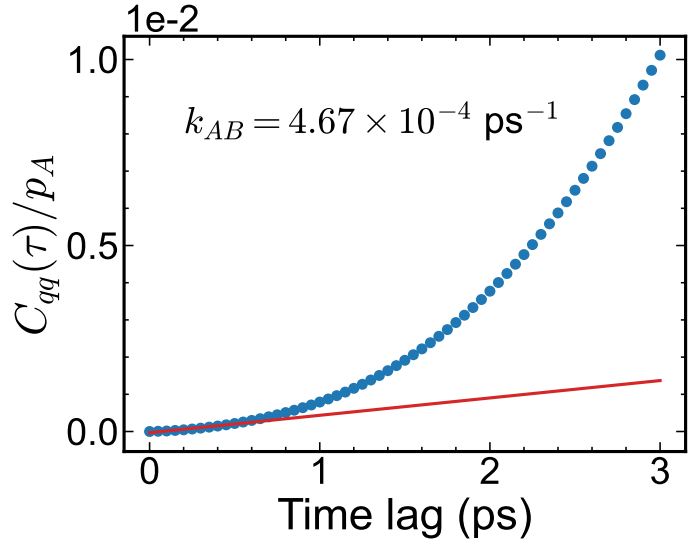

**Fig. 13**: Rate constant $k_{AB}$ determined from PCCANN using the biased trajectory of the one-dimensional double-well potential.

## Berezhkovskii-Szabo Two-Dimensional Potential

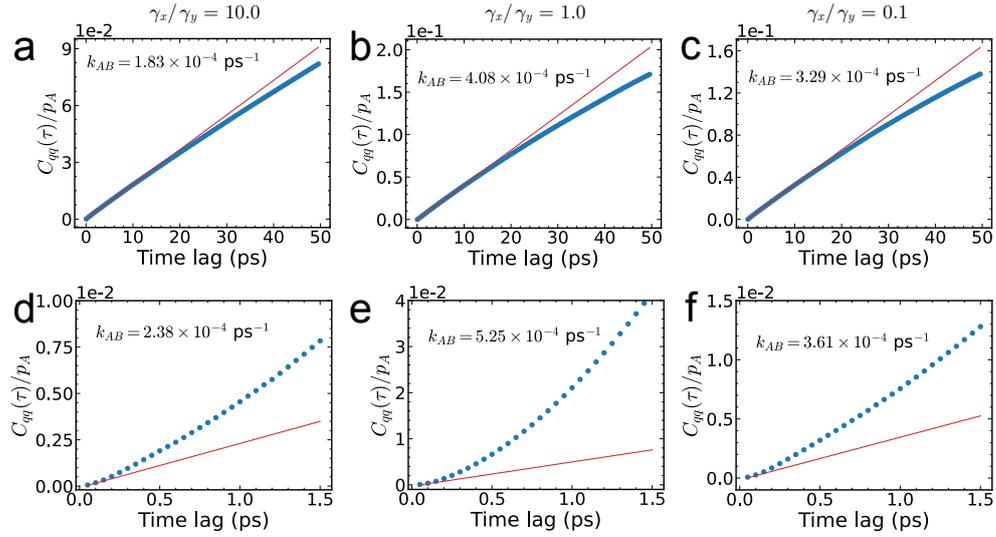

**Fig. 14**: Rate constant $k_{AB}$ determined from PCCANN using (a-c) unbiased and (d-f) biased trajectories of the BS potential. Each column corresponds to a diffusion value specified at the top of the corresponding column.

19

## Triple-well Potential

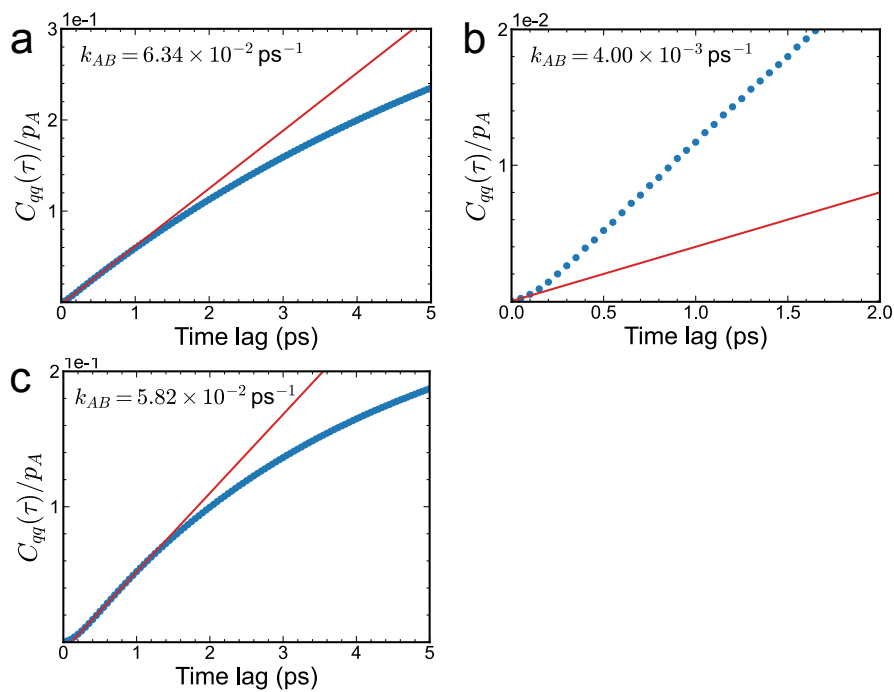

**Fig. 15**: Rate constant $k_{AB}$ determined from PCCANN using unbiased trajectory of the triple-well potential at (a) 300 K, and biased trajectories at (b) 100 K and (c) 300 K, along the upper and lower channels respectively.

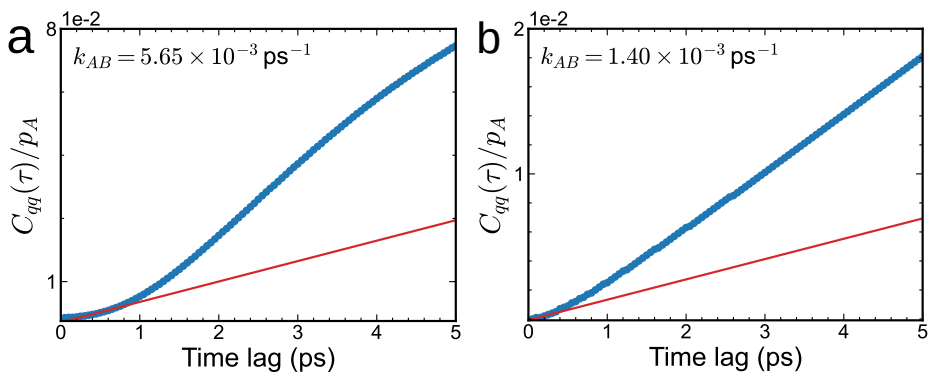

**Fig. 16**: Rate constant $k_{AB}$ determined from PCCANN using biased trajectories at 300 K along the (a) upper and (b) lower reactive trajectories for a very narrow tube covering the string.

## Müller-Brown Potential

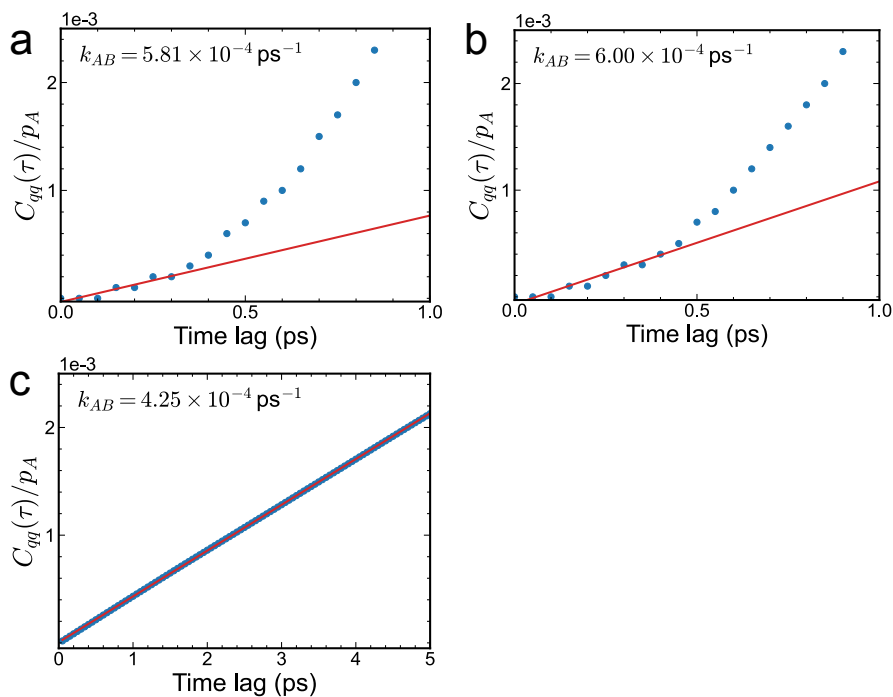

**Fig. 17**: Rate constant $k_{AB}$ determined from PCCANN using (a) unbiased and (b) biased trajectory of the MB two-dimensional potential. (c) Rate constant $k_{AB}$ determined from PCCANN of the MB potential on a three-dimensional Swiss roll structure.

## $N$–Acetyl–$N'$–methylalanylamide

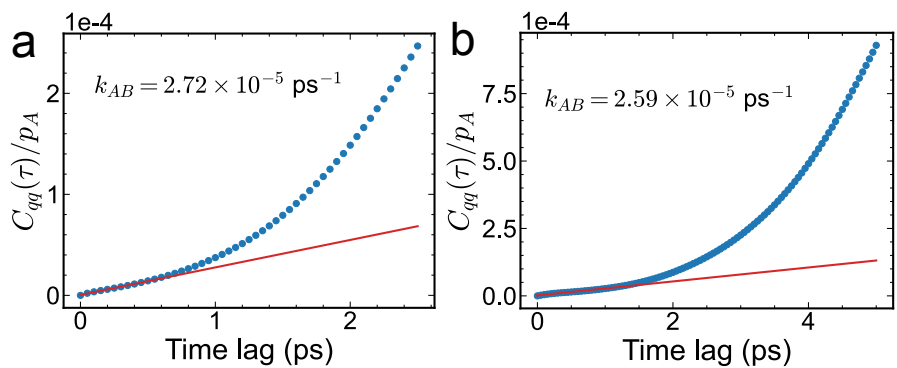

**Fig. 18**: Rate constant $k_{AB}$ determined from PCCANN using biased simulations for (a) $(\phi, \psi)$ and (b) $(\phi, \psi, \theta, \omega)$ as the CV of NANMA.

21

## Chignolin

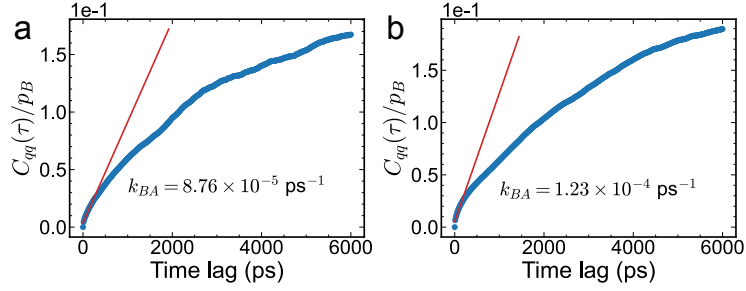

**Fig. 19**: Rate constant $k_{AB}$ determined from PCCANN using a biased trajectory along the first (a) and second path (b) of chignolin.

| | Transition rate, $k_{AB}$ (ps$^{-1}$) | | | Pontryagin et al.[28] | |
|---|---|---|---|---|---|
| | MFPT | PCCANN | | | |
| System | Unbiased | Unbiased | Biased | SMwST | CCS |
| **Double-well potential** | | | | | |
| | $4.93 \times 10^{-4}$ | | $4.67 \times 10^{-4}$ | — | — |
| **Berezhkovskii-Sazbo potential**[1] | | | | | |
| $\gamma_x/\gamma_y = 0.1$ | $3.36 \times 10^{-3}$ | $3.29 \times 10^{-3}$ | $3.61 \times 10^{-3}$ | — | — |
| $\gamma_x/\gamma_y = 1.0$ | $4.27 \times 10^{-3}$ | $4.08 \times 10^{-3}$ | $5.24 \times 10^{-3}$ | — | — |
| $\gamma_x/\gamma_y = 10.0$ | $1.86 \times 10^{-3}$ | $1.83 \times 10^{-3}$ | $2.38 \times 10^{-3}$ | — | — |
| **Müller-Brown potential**[1,34] | | | | | |
| Two dimensional | $5.37 \times 10^{-4}$ | $5.81 \times 10^{-4}$ | $6.00 \times 10^{-4}$ | $3.38 \times 10^{-4}$ | $2.94 \times 10^{-4}$ |
| Swiss roll | — | $4.25 \times 10^{-4}$ | — | — | — |
| **Triple-well potential**[15,35] | | | | | |
| 100 K | $5.70 \times 10^{-5,a}$ | — | $4.00 \times 10^{-3}$ | $2.95 \times 10^{-3}$ | $2.01 \times 10^{-3}$ |
| 300 K lower path$^b$ | $6.44 \times 10^{-2}$ | $6.34 \times 10^{-2}$ | $5.82 \times 10^{-2}$ | $5.18 \times 10^{-2}$ | $4.66 \times 10^{-2}$ |
| 300 K lower path$^c$ | $6.44 \times 10^{-2}$ | $6.34 \times 10^{-2}$ | $5.65 \times 10^{-3}$ | $5.18 \times 10^{-2}$ | $4.66 \times 10^{-2}$ |
| 300 K upper path$^c$ | $6.44 \times 10^{-2}$ | $6.34 \times 10^{-2}$ | $1.40 \times 10^{-3}$ | — | — |
| **N–Acetyl–N'–methylalanylamide**[36] | | | | | |
| $\phi, \psi$ | $1.34 \times 10^{-5}$ | — | $2.72 \times 10^{-5}$ | $4.99 \times 10^{-6}$ | $5.40 \times 10^{-6}$ |
| $\phi, \psi, \theta, \omega$ | $1.34 \times 10^{-5}$ | — | $2.59 \times 10^{-5}$ | $3.30 \times 10^{-6}$ | $3.29 \times 10^{-6}$ |
| **Chignolin** | | | | | |
| 1st path | $1.6 \times 10^{-6}$ [37] | — | $8.76 \times 10^{-5}$ | — | — |
| 2nd path | | — | $1.23 \times 10^{-4}$ | — | — |

$^a$This value is computed at a low temperature, where very rare transitions occur.
$^b$Simulations performed on a very wide tube embracing the string.
$^c$Simulations performed on a very narrow tube embracing the string.

**Table 1**: Summary of the transitions rates $k_{AB}$ for the one-dimensional double-well (DW) potential, the two-dimensional Berezhkovskii-Sazbo (BS) potential, the two-dimensional Müller-Brown (MB) potential, the two-dimensional triple-well (TW) potential, the $N$–Acetyl–$N'$–methylalanylamide (NANMA), and the mini-protein chignolin. The transition rates, $k_{AB}$, are determined using mean first-passage times (MFPT) from unbiased simulations, the PCCANN from unbiased and biased simulations, and from the integral 17 proposed by Pontryagin et al[28,29]. For chignolin, $k_{BA}$ was estimated instead, in order to make the comparison with the value reported in reference 37.


\end{document}